\documentclass[letterpaper,twocolumn,10pt]{article}
\usepackage{usenix-2020-09}

\usepackage{microtype}
\usepackage{graphicx}
\usepackage{subcaption}
\usepackage{booktabs} % for professional tables

\usepackage{amsmath}
\usepackage{amssymb}
\usepackage{comment}
\usepackage{dsfont}
\usepackage{graphicx}
\usepackage{xfrac}
\usepackage{extarrows}
\usepackage{amsbsy}
\usepackage[utf8]{inputenc}
\usepackage{listings,xcolor}
\usepackage[T1]{fontenc}
\usepackage{xcolor}
\usepackage[scaled=0.9]{DejaVuSansMono}
\usepackage{tablefootnote}
\usepackage{array}
\usepackage[toc,page]{appendix}
\usepackage{booktabs}
\usepackage{makecell}
\usepackage{placeins}
\usepackage[compress]{cleveref}
\usepackage{xspace}
\usepackage{algorithm}
\usepackage{algpseudocode}
\usepackage{multirow}
\usepackage{pifont}
\usepackage[normalem]{ulem}
\usepackage[switch]{lineno}
\usepackage{url}

\definecolor{commentcolor}{HTML}{9596ad}
\lstdefinestyle{mypython}{
    language=Python,
    basicstyle=\small\ttfamily,
    keywordstyle=\bfseries,
    commentstyle=\color{commentcolor}\itshape,
    columns=fullflexible,
    keepspaces=true,
    breaklines=true,
    showstringspaces=false,
    lineskip={-1pt},
    morekeywords={maximum, argmin, argmax, items}
}
\makeatletter
\newenvironment{codealgorithm}[1][htb]{%
  \renewcommand{\ALG@name}{Algorithm}%
  \begin{algorithm}[#1]%
  }{\end{algorithm}}
\makeatother

\def\varequals#1{\mathrel{\leaders\hbox to3pt{$=$\hss}\hskip#1=}} % equation equal sign

\newtoggle{showmarks}
\toggletrue{showmarks}

\iftoggle{showmarks}{
    \newcommand\plan[1]{\textcolor{gray}{#1}}
    \newcommand\todo[1]{\textcolor{blue}{TODO: #1}}
    \newcommand\plot[1]{\hfill\newline\textcolor{blue}{[plot] #1\newline}}
    \newcommand\askiad[1]{\textcolor{cyan}{[Athinagoras:] #1}}
    \newcommand\myzhao[1]{\textcolor{purple}{[Mark: #1]}}
    \newcommand\shm[1]{\textcolor{orange}{[Shrijeet:] #1}}
    \newcommand\christos[1]{\textcolor{magenta}{[Christos:] #1}}
    \newcommand\thomas[1]{\textcolor{violet}{[Thomas: #1]}}
    \newcommand\old[1]{\textcolor{gray}{#1}}
    
    \newcommand\rem[1]{\textcolor{pink}{\sout{#1}}}
}{
    \newcommand\plan[1]{\unskip}
    \newcommand\todo[1]{\unskip}
    \newcommand\plot[1]{\unskip}
    \newcommand\askiad[1]{\unskip}
    \newcommand\myzhao[1]{\unskip}
    \newcommand\shm[1]{\unskip}
    \newcommand\christos[1]{\unskip}
    \newcommand\thomas[1]{\unskip}
    \newcommand\old[1]{\unskip}
    
    \newcommand\rem[1]{\unskip}
}

\newcommand{\NAME}{\textsc{Symi}\xspace} % please fix this

\usepackage{tikz}% tikz
\DeclareRobustCommand\numcircledtikz[1]{\tikz[baseline=(char.base)]{
    \node[shape=circle,draw,fill,inner sep=1pt] (char)
    {\textcolor{white}{#1}};}}
\begin{document}
%-------------------------------------------------------------------------------

%don't want date printed
\date{}

\title{\NAME: Efficient Mixture-of-Experts Training \\
via Model and Optimizer State Decoupling}

%for single author (just remove % characters)
% \author{
% {Paper \# 412}
% } % end author

\author{
{\rm Athinagoras Skiadopoulos}\\
Stanford University\\
{\rm askiad@stanford.edu}\\
\and
{\rm Mark Zhao}\\
University of Colorado Boulder\\
{\rm myzhao@colorado.edu}
\and
{\rm Swapnil Gandhi}\\
Stanford University \& NVIDIA\\
{\rm gandhis@stanford.edu}
\and
{\rm Thomas Norrie\NoHyper\thanks{Work done while at Enfabrica.}\endNoHyper}\\
OpenAI\\
{\rm tnorrie@openai.com}
\and
{\rm Shrijeet Mukherjee\footnotemark[1]}\\
NVIDIA\\
{\rm shrijeetm@nvidia.com}
\and
{\rm Christos Kozyrakis}\\
Stanford University \& NVIDIA\\
{\rm kozyraki@stanford.edu}
} % end author

\maketitle
% \linenumbers
% \pagestyle{empty}

\begin{abstract}
%\christos{edit abstract towards the end, roughly 1 sentence per paragraph in intro}
% The continued scaling of large language models (LLMs) has unlocked unparalleled performance across a variety of tasks.
% This scaling requires hardware resources to scale accordingly, hitting the compute and memory limits of modern training infrastructures. 
% To bypass the compute wall, LLMs are increasingly relying on sparse Mixture-of-Experts (MoE) architectures that allow continued parameter scaling with sub-linear compute cost. % need to shorten the above

%The continued scaling of large language models (LLMs) has unlocked unparalleled performance across a variety of tasks.
Mixture-of-Experts (MoE) models have become a widely-adopted solution to continue scaling model sizes without a corresponding linear increase in compute. During MoE model training, each input token is dynamically routed to a subset of experts -- sparsely-activated feed-forward networks -- within each transformer layer. The distribution of tokens assigned to each expert varies widely and rapidly over the course of training. To handle the wide load imbalance across experts, current systems are forced to either drop tokens assigned to popular experts, degrading convergence, or frequently rebalance resources allocated to each expert based on popularity, incurring high state migration overheads.

To break this performance-accuracy tradeoff, we introduce \NAME, an adaptive MoE training system. The key insight of \NAME is to decouple the placement of expert parameters from their large optimizer state. \NAME statically partitions the optimizer of each expert across all training nodes. Meanwhile, \NAME dynamically adjusts the placement of expert parameters by repurposing existing weight updates, avoiding migration overheads. In doing so, \NAME right-sizes the GPU resources allocated to each expert, on a \textit{per-iteration} basis, with minimal overhead. Compared to state-of-the-art MoE training systems, DeepSpeed and FlexMoE, \NAME is able to achieve a $30.5\%$ and $25.9\%$ faster time-to-convergence, respectively.

\end{abstract}

% traditionally, model state and the rest (optimizer) have been coupled
% we break this 
% allow then to be designed and optimized independently
% in this study, we allow the model state (experts) move dynamically and be replicated adaptively for maximum convergence
% we keep the optimizer (decoupled from the model) static, for minimum movement and system overhead
% our design allows future innovation by new future directions (data management, sharding, placement, topology-aware model architectures)
\section{Introduction}
\label{sec:introduction}

 Recent breakthroughs in large language models (LLMs) have been largely unlocked by {\it massively scaling} their size, i.e., the number of parameters in the model~\cite{kaplan2020scaling}. 
State-of-the-art foundation models now consist of hundreds of billions to trillions of parameters~\cite{arxiv:smith_turingnlg, arxiv:chowdhery_palm, arxiv:ren_pangu, GPT4, llama4, grok, website:gemini, liu2024deepseek, mixstral, snowflakeArctic, yang2025qwen3}.
Training such models already requires several weeks of compute on systems with thousands of GPUs -- further scaling model size without architectural improvements is unsustainable.

Sparse architectures, which selectively activate parameters, are rapidly growing in popularity due to their compute efficiency.
Most frontier models today, including GPT4~\cite{GPT4}, Llama 4~\cite{llama4}, Gemini 2.5~\cite{website:gemini}, Grok-1~\cite{grok}, Qwen3~\cite{yang2025qwen3},
% Mistral AI's Mixtral~\cite{mixstral}, 
and DeepSeek-R1~\cite{guo2025deepseek} use the Mixture-of-Experts (MoE) architecture~\cite{jmlr22:fedus_switch-transformers}.
MoE models decouple parameter scaling from compute scaling by using each input token to train only a subset of parameters per layer (i.e., experts).
Recent MoE models feature 8 to 512 experts per layer~\cite{pmlr22:rajbhandari_deepspeed-moe, liu2024deepseek, arxiv:jiang_mixtral, website:dbrx, arxiv:du_glam, llama4, mlsys23:hwang_tutel, qwen3-next}.
Increasing the number of experts enables model size scaling with sublinear increase in compute demands.

To train MoE models, a learned router dynamically assigns input tokens to experts in each iteration.
The distribution of tokens assigned to experts is highly dynamic, varying \textit{widely} and \textit{rapidly} across training iterations.
As shown in Figure~\ref{fig:motivation_distr}~(\S\ref{sec:background_moe}), the number of tokens routed to each expert can fluctuate by over $16\times$ in as few as 3 iterations.
Thus, MoE training introduces a provisioning challenge.
Letting experts process their full token load increases iteration latency, as popular experts become bottlenecks while less popular experts remain idle. 
Conversely, capping each expert's token processing capacity degrades convergence, as popular experts must drop excess tokens~\cite{zhou2022choice, Zoph2022-stmoe, Yang2021-m6t, jmlr22:fedus_switch-transformers, mlsys23:hwang_tutel}.
A non-uniform but static provisioning among experts is insufficient as expert popularity varies widely during training.

To break this {\it performance-accuracy tradeoff}, an ideal system would 
% assign GPU resources according to the popularity of each expert, by dynamically adjusting each expert's \textit{replication} factor.
replicate experts non-uniformly and {\it adaptively} to their {\it dynamic popularity}.
This would minimize token drops for better convergence without performance penalty.
However, adaptive expert replication introduces major system challenges.
Expert rebalancing incurs the heavy overhead of transferring expert parameters and optimizer states across GPUs. 
As a result, current systems that support adaptive expert replication limit rebalancing frequency, e.g., every 50-100 iterations~\cite{atc23:zhai_smartmoe, sigmod23:nie_flexmoe, yu2024moesys}, thereby limiting the performance and accuracy benefits of adaptive replication.

The optimizer state in particular is massive.
For example, the Adam optimizer of an 8B model consumes 96.6GB (excluding gradients)~\cite{badam2024nips}, exceeding an H100’s HBM~\cite{nvidiaH100}.
In modern MoE models, individual experts often surpass this 8B mark~\cite{zhou2023brainformers, arxiv:du_glam, mixstral, llama4, grok, website:dbrx, zhou2022choice}.
In the trillion-parameter regime, experts scale further, commensurately amplifying the per-expert optimizer footprint.
Consequently, optimizer sharding and/or offloading for both the dense and sparse model components are widely used and essential for training large models~\cite{rasley2020deepspeed, sc21:rajbhandari_zero-infinity, singh2024study, liu2024deepseek, zhao2023pytorch, zhang2022opt}.
Our work is situated within this context of optimizer sharding for large MoE models.

In this paper, we show that expert replication can be rebalanced on every iteration without extra data movement compared to normal training.
Our key insight is to \textit{decouple the adaptively-replicated expert parameters from their optimizer state}.
We realize this insight in a system called \NAME to achieve efficient and frequent adaptive expert replication.
\NAME evenly and statically partitions each expert's optimizer state across host memory, independently from the replication scheme of expert parameters in GPU memory. \NAME never relocates the optimizer state but adjusts experts' replication proportionally to their popularity on a \textit{per-iteration basis}.
\NAME reshuffles expert weights by repurposing existing communication during the optimizer pass.
Whether a GPU receives the updated weights of a previous or a newly assigned expert, data volume is the same.
Hence, \NAME does not introduce any extra data movement during expert rebalancing.

% \NAME's adaptive replication approach exhibits new communication patterns for expert rebalancing as it transfers gradients and weights between the dynamic experts and the static optimizer. We introduce new communication collectives to rebalance experts, replacing existing communication performed in static systems.
% While the overall communication volume remains invariant, \NAME alters expert-to-optimizer locality.
% \NAME's collectives reduce communication cost by implementing locality-enhanced expert placement (favoring the placement of same-class expert replicas within the same rank), and load-balanced gradient aggregation (avoiding bottlenecks during gradient communication to the optimizer). 
% % \christos{can you add a sentence that say that the locality enchanced gradient sync and that the load balanced gradient aggregation are?}
% \NAME then materializes each iteration's expert placement by dynamically adjusting the ranks receiving the updated optimizer weights.

\NAME's adaptive replication exhibits new communication patterns transferring gradients and weights between dynamic experts and the static optimizer. We introduce new collectives to rebalance experts, replacing existing communication performed in static systems.
While the overall communication volume remains invariant, \NAME alters expert-to-optimizer locality.
\NAME's collectives reduce communication cost by implementing locality-enhanced expert placement (favoring placement of same-class expert replicas within the same rank), and load-balanced gradient aggregation (avoiding bottlenecks during gradient communication to the optimizer). 
\NAME then materializes each iteration's expert placement by dynamically adjusting the ranks receiving the updated optimizer weights.

We implemented \NAME on top of DeepSpeed, demonstrating $30.5\%$ and $25.9\%$ faster convergence compared to DeepSpeed and coarse-grained adaptive expert replication solutions, respectively. These gains stem from \NAME dropping $43\%$–$69\%$ fewer tokens than compared systems, while incurring no additional overhead relative to DeepSpeed. In contrast, existing adaptive expert replication methods increase iteration latency by $2.46\times$–$4.10\times$ during rebalancing.
%\christos{expland on this tracking what you have in evaluation. We show X benefit over static and Y benefit over coarser-grain dynamic. The benefits stem from ... with numbers}

\section{Background and Motivation}
\label{sec:background}
\subsection{Mixture-of-Experts}
\label{sec:background_moe}
\begin{figure}[t]
  \centering
  \includegraphics[width=\linewidth]{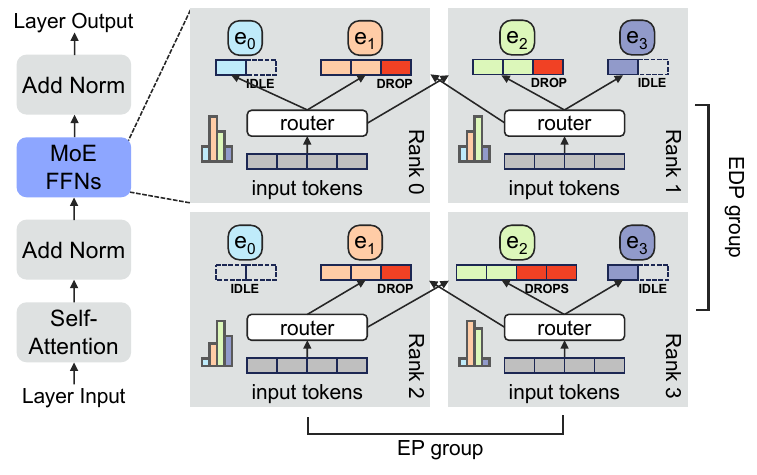}
  \caption{Overview of an MoE layer and expert parallelism.}
  \label{fig:bgrd_moe}
\end{figure}

\noindent\textbf{MoE Architecture.} To enable continued parameter scaling without a corresponding increase in compute requirements, many modern LLMs~\cite{llama4, GPT4, website:gemini, grok, mixstral, liu2024deepseek} rely on an MoE architecture.
As shown in Figure~\ref{fig:bgrd_moe}, MoE architectures are built upon a traditional Transformer architecture~\cite{neurips17:vaswani_transformer}, except that the traditional dense feed-forward network (FFN) in each layer is replaced with a number of \textit{experts}.
Each expert has the same dimensions as the original FFN, but is trained independently, increasing the total number of trainable parameters of the model.
Each layer routes each input token to a subset of experts, allowing each expert to specialize for a distinct portion of the input space~\cite{jmlr22:fedus_switch-transformers}.
Modern LLMs rely on a large and increasing number of experts, ranging from 8 to upwards of 512 of experts per layer~\cite{pmlr22:rajbhandari_deepspeed-moe, liu2024deepseek, arxiv:jiang_mixtral, website:dbrx, arxiv:du_glam, llama4, mlsys23:hwang_tutel, qwen3-next}.

% ~\cite{pmlr22:rajbhandari_deepspeed-moe, arxiv:jiang_mixtral, website:dbrx, jmlr22:fedus_switch-transformers, arxiv:du_glam, sigmod23:nie_flexmoe, mlsys23:hwang_tutel, .

%!TEX root =paper.tex

As shown in Figure~\ref{fig:bgrd_moe}, each MoE layer selects the expert(s) to activate for each token via a learned \textit{router} or \textit{gate network}.
The router receives the input token embeddings from the layer's attention block and selects the Top-$k$ experts for each token.
The selected experts are activated and process their assigned tokens.
Notably, for $k=1$ (each token is routed to a single expert), the sparsely-activated MoE model performs similar number of FLOPS per token compared to the respective dense model.
% \myzhao{"equivalent" may be a bit strong due to minor differences (e.g., router computation). Maybe "similar" or "comparable"?}
% \footnote{We assume a Top-1 selection~\cite{jmlr22:fedus_switch-transformers} for the remainder of this paper.}. \christos{you can actually dro that whole Notably sentence and nothign bad will happen. Avoid the possibiliy of confusion}

\noindent\textbf{Expert Parallelism.}
To avoid the memory overhead of allocating a copy of every expert on every rank, systems use expert parallelism~\cite{jmlr22:fedus_switch-transformers, arxiv:lepikhin_gshard, arxiv:shazeer_outrageously}.
As illustrated in Figure~\ref{fig:bgrd_moe}, expert parallelism distributes experts across ranks (i.e., GPUs).
Each rank hosts a fixed number of \textit{expert slots}, 
and each slot is assigned an \textit{expert class}.
The set of ranks hosting the different expert classes form an expert parallel (EP) partition.
Scaling out, each expert class is replicated an equal number of times with expert data parallelism (EDP)~\cite{ics23:singh_deepspeed-ted}.
We refer to the sum of all replicas as \textit{expert instances}.

Because expert instances are distributed across ranks within each EP partition, two all-to-all collective communications are needed in the forward pass to scatter input tokens and gather expert outputs, and two all-to-all collectives are needed in the backward pass to scatter and gather gradients.
Furthermore, an all-reduce collective is needed within each EDP partition in the backward pass to synchronize gradients across expert instances of the same expert class, as with data parallelism.

% Effect of capacity to convergence vs latency - load balance to resolve tradeoff
\noindent\textbf{Dynamic Expert Activation.}
% !TEX root =paper.tex
\begin{figure}[t]
  \centering
  \includegraphics[width=\linewidth]{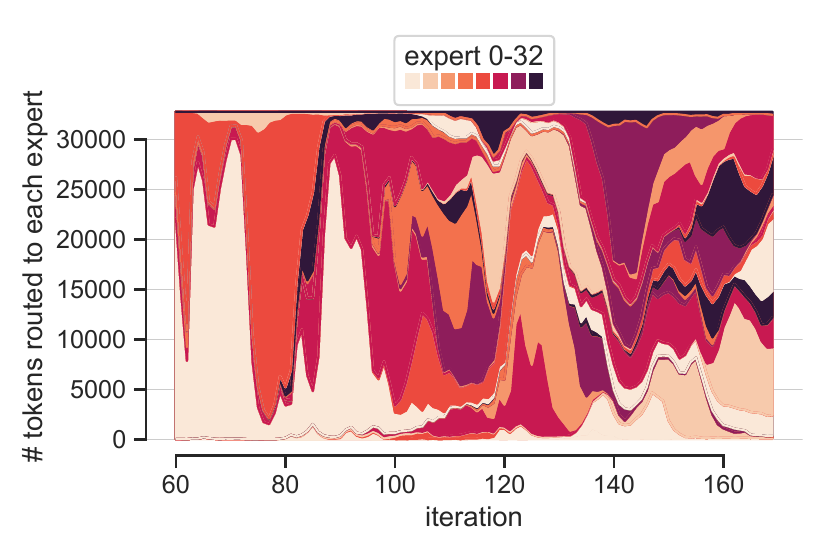}
  \caption{
  % \new{
  A single layer's % }
  expert popularity distribution during the training of GPT-Small (125M) extended with 32 experts.
  % \rem{per layer}. 
  The distribution shifts dramatically within very few iterations.
  % \christos{add a sentence about which model is this or a forward pointer to the methodology section}
  }
  \label{fig:motivation_distr}
\end{figure}
The router dynamically assigns tokens to expert classes on each iteration.
As shown in Figure~\ref{fig:motivation_distr}, the token distribution across experts can be both \textit{highly skewed} -- different expert classes receive disproportionately more tokens, and \textit{highly dynamic} -- varying significantly, even within a few iterations~\cite{jmlr22:fedus_switch-transformers, sigmod23:nie_flexmoe, atc23:zhai_smartmoe, atc23:li_lina, ppopp22:he_fastermoe, liu2024deepseek, mlsys23:hwang_tutel, sigcomm23:liu_janus, ppopp25:wang_ccfuser, sc24:wei_aptmoe}.
Figure~\ref{fig:motivation_distr} shows cases where 
expert token load fluctuates by over $16\times$ within only 3 iterations (e.g., iterations 72-75).
% \christos{can you draw attention to soem specific iteraiton, eg, around iteration 120}
%  or reduces by $9.2\times$ 

\noindent\textbf{The Convergence-Latency Tradeoff:} In traditional expert parallelism with uniform replication, popular experts become latency bottlenecks~\cite{sigmod23:nie_flexmoe, atc23:zhai_smartmoe, ppopp22:he_fastermoe}.
Specifically, devices hosting popular experts delay iteration completion as they process more tokens while other devices idle.
Additionally, devices with popular experts become a bottleneck in the all-to-all collectives, receiving more tokens and sending more activations.

% !TEX root =paper.tex
\begin{table}[t]
    \centering
        \caption{The convergence-latency tradeoff for different expert capacities. Shown is GPT-Small (125M) extended with 32 experts per layer, ran in a 16 GPU cluster.}
    \begin{tabular}{c c c c}
    \toprule
    \makecell{\textbf{Expert}\\ \textbf{Capacity}} &
    \makecell{\textbf{Avg. Token}\\ \textbf{Survival (\%)}} &
    \makecell{\textbf{Iters to}\\ \textbf{Target Loss}} &
    \makecell{\textbf{Forward Pass}\\ \textbf{Latency (ms)}} \\
    \midrule
    $\times1$ & 44.90 & 618 & 455.41 \\
    $\times2$ & 65.56 & 527 & 506.77 \\
    $\times4$ & 74.91 & 478 & 571.42 \\
    \bottomrule
    \end{tabular}
\label{tab:motiv_cap}
\end{table}

% \christos{make it clear that in practice everybody sets a capacity (and drops, they don't just way as discussed in prev paragraph. } 
To manage this load imbalance, the standard practice in today's MoE systems is setting an expert \textit{capacity}, which defines the maximum number of tokens each expert class can process~\cite{jmlr22:fedus_switch-transformers}.
Tokens that exceed this expert capacity are dropped, improving system latency.
However, dropping tokens results in slower convergence and model quality degradation~\cite{jmlr22:fedus_switch-transformers, zhou2022choice, Zoph2022-stmoe, Yang2021-m6t, mlsys23:hwang_tutel, cai2025survey}.
Table~\ref{tab:motiv_cap} demonstrates this tradeoff:
reducing expert capacity yields ${\sim}20\%$ latency improvement, especially critical for modern AI systems that train models over thousands of GPUs.
However, higher capacity increases the token survival rate (by ${\sim}30$ percentage points) which is shown to be directly correlated with convergence speed.
% \myzhao{I would add a sentence discussing the actual results. Also, if you show the full iteration time or higher expert capacity would there be a more drastic difference in latency?}
Consequently, MoE training systems have an intrinsic tradeoff between model convergence and system performance.
% \christos{A reader may not be able to tell how big of a deal is 5.43 vs 5.26 in training loss. Similarly, the fwd pass latency seems small. So add a couple of sentences discussing why these results are big/important, etc}

% \new{
On top of expert capacity, a number of techniques have been introduced to alleviate the token load imbalance~\cite{arxiv:lepikhin_gshard, mlsys23:gale_megablocks, wang2024auxiliary, zhou2022choice}. %(Section~\ref{sec:related}).
For example, an auxiliary load-balancing loss is commonly applied to penalize uneven expert utilization, thereby reducing token drops and latency.
Its coefficient, however, requires careful tuning. 
High auxiliary loss involvement harms convergence; it can overwhelm the main loss objective and steer experts towards sub-optimal specialization~\cite{jmlr22:fedus_switch-transformers, sigmod23:nie_flexmoe, wang2024auxiliary, Zoph2022-stmoe, zhou2022choice, guo2025advancing, dai2024deepseekmoe, omi2025simbal},
further highlighting the convergence-latency tradeoff.
% }

% \todo{add concrete numbers here on extra latency for high capacity, and convergence degradation for low capacity}
% \askiad{I've run the experiment and makes sense to add. Will add a convergence - drop - latency comparison for different capacity factors.}

\subsection{Adaptive Expert Replication}
\label{sec:background_adaptive}

\noindent\textbf{Dynamic, Load-Aware Expert Replication.}
The root cause of the convergence-latency tradeoff is that expert replication is uniform and static, while expert popularity is skewed and dynamic.
An adaptive expert replication strategy can address this issue by adjusting each expert's replication degree dynamically, in proportion to their popularity.
In the ideal case where replication precisely matches popularity, tokens can be routed to their assigned expert classes with minimum iteration latency and minimum drop rate.
Popular expert classes receive sufficient replicas to handle their token load without extra overhead or token drops, while less popular expert classes are allocated fewer replicas to avoid idle time.

\noindent\textbf{Rebalancing Cost.}
Unfortunately, adaptive expert replication comes with major system challenges.
Every expert rebalancing produces high overhead due to a blocking shuffle required to redistribute the expert's state across ranks. 
%state redistribution, primarily due to migrating the expert's optimizer state.
%The primary source of this cost is the migration of optimizer state.
% The principal problem with adaptive expert replication is the high cost incurred every time experts are rebalanced (i.e., modifying replication degrees to adjust to the evolving popularity).
%Every rebalancing requires a blocking shuffle of expert state across ranks.
Specifically, assigning a new expert to a slot requires moving both the expert's weights (2B per parameter) and \textit{optimizer state} (16B per parameter)\footnote{We assume fp16/fp32 mixed precision training with the Adam optimizer.}~\cite{sc21:rajbhandari_zero-infinity} across GPUs.
For example, for a typical model dimension of 12288 as in GPT3-175B~\cite{brown2020language, sc21:narayanan_training-megatron}, and a state-of-the-art 400 Gbps GPU-to-GPU InfiniBand interconnect,
rebalancing a \textit{single} expert within a \textit{single} layer would require 
transferring $3.375$GB of model weights and $27$GB of optimizer state~\cite{sc21:rajbhandari_zero-infinity}, incurring a latency of $0.0675$s and $0.54$s, respectively. 
Moving a single expert's optimizer in particular is on par with typical iteration latencies of $1$s~\cite{sc21:narayanan_training-megatron}.
% \christos{a reader may wonder why not this within NVLink so you can transfer at higher bandwidth. may need a comment about scaleout vs up somewhere (later?)}
Given the highly dynamic nature of expert popularity, frequently rebalancing experts to align with shifting popularities incurs overheads impractical for current systems.

%rebalancing rebalancing costs frequently renders adaptive replication impractical in current systems.

% \christos{if FlexMOE is the major comparison point in the results, you have to explain it here (and name it). Say that it does a couple of things, adapt the frequency and rebalance only a few experts. But as yo ushow later not enough} 

\noindent\textbf{Current Approaches.}
Modern MoE frameworks employ ad-hoc solutions to sidestep this challenge. 
Existing adaptive expert replication systems rebalance experts infrequently based on hard-coded thresholds, use heuristics, and rebalance only a subset of experts per update~\cite{sigmod23:nie_flexmoe, atc23:zhai_smartmoe, yu2024moesys}.
FlexMoE~\cite{sigmod23:nie_flexmoe} triggers expert rebalancing based on a predefined popularity skewness threshold (equivalent to every 50-100 iterations)
and iteratively shifts by one replica the most and least popular experts until another cost-based threshold is crossed.

These approaches cannot handle the frequent variation in expert distributions shown in Figure~\ref{fig:motivation_distr}, causing them to leave accuracy and performance benefits on the table as with static replication mechanisms.
%As we show in \S\ref{sec:eval_convergence}, our proposed system achieves $25.9\%$-$29.4\%$ faster time to convergence than coarser-grained adaptive replication solutions.
% \myzhao{a quick forward pointer to show magnitude of the problem for perf/accuracy in current systems.}
An ideal system should enable \textit{fine-grained adaptive replication}, at each iteration, while doing so \textit{efficiently}, without overheads from moving expert state.

\section{\NAME Design}
\label{sec:design}

% \christos{When sec 1 and 2 are in we need to make another pass for consistency of messaging}

% Summary: Expert placement changes per iteration. Symi optimizer is static -- offloaded and uniformly sharded across all ranks for all experts. In optimizer step; (1) gather gradients from experts (experts follow current iteration placement), (2) compute the updated weights in the CPU, (3) scatter expert model weights to GPUs using the new distribution.

\subsection{Key Insight}

The key challenge in dynamically and frequently rebalancing experts is managing the overhead of moving expert weights, and more critically, the large optimizer state -- traditionally tied to its corresponding weights. We design \NAME to eliminate this overhead entirely.

% !TEX root =paper.tex
\begin{figure}[t]
  \centering
  \includegraphics[width=\linewidth]{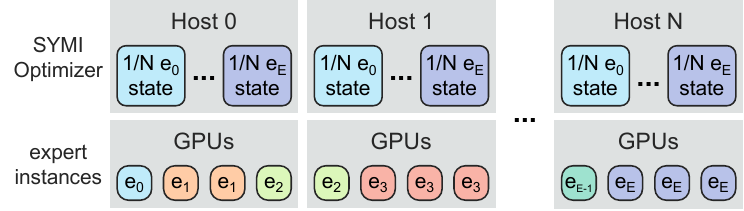}
  \caption{\NAME decouples the model and optimizer state placement. Expert replication is non-uniform and dynamic, while each expert's optimizer remains static, uniformly sharded across all hosts.}
  \label{fig:design_decouple}
\end{figure}

Our key insight is to \textbf{decouple the experts optimizer state from expert instances}, allowing the optimizer offloading, sharding, and placement to be independent of the replication and placement of its corresponding expert instance(s).
% \new{
% % As model size grows, the optimizer state (multiple TBs for trillion-parameter models) exceeds the capacity of a single device/node.
% As model size grows, the optimizer state reaches multiple terabytes for trillion-parameter models, exceeding the capacity of a single device/node.
% Optimizer sharding and offloading are widely used and crucial for enabling larger models~\cite{rasley2020deepspeed, sc21:rajbhandari_zero-infinity, singh2024study, liu2024deepseek, zhao2023pytorch, zhang2022opt}.
% }
As shown in Figure~\ref{fig:design_decouple}, \NAME offloads\footnote{
% \new{
Offloading the optimizer cleanly separates static from dynamic components in \NAME. Still, our design principles are independent of the memory tier where the optimizer is placed (see Section~\ref{sec:discussion}).
% }
} and statically shards the optimizer state for each expert uniformly across all nodes, regardless of expert placement. Meanwhile, \NAME replicates experts non-uniformly and dynamically, right-sizing replication according to popularity.

% For example, in Figure~\ref{fig:strawman} \textit{each} rank has an equal partition (1/256-th) of the optimizer state for \textit{all} 64 experts.
% Figure~\ref{fig:strawman} also shows why doing so meets all three key requirements (adaptive replication, efficient rebalancing, and efficient memory management).

\noindent\textbf{No-overhead adaptive replication.}
In contrast to existing adaptive replication solutions, \NAME \textit{never moves the optimizer state}.
This avoids the immense cost of migrating optimizer state with each rebalancing.
While optimizer state is static, model state (expert instances) is dynamic.
At the end of each iteration, the optimizer always needs to perform necessary communication to move the updated expert weights to their corresponding instances in GPUs.
\NAME makes the critical observation that this data movement volume is invariant, regardless of the actual data content corresponding to different expert assignments.
The optimizer can choose to transfer to any given expert slot either the updated weights of the previously assigned expert or those of a newly assigned one.
As a result, \NAME enables shuffling into a new, arbitrary expert placement after each optimizer step, \textit{without requiring any additional data movement}. 
% \christos{I would start with some of the observations in this paragraph earlier in the section. Something like: In Symi, we make the following critical observation: the challenge in dynamically rebalancing experts is the moving optimizer state; model weights always need to be updated across the system at the end of each iteration. Based  on that, we design Symi... You can have more details in the following paragraphs of course} \askiad{How is it now?}

\noindent\textbf{Continuous adaptive replication.}
\NAME \textit{replicates experts proportionally to their popularity}, thus minimizing token drops and improving model convergence.
% \shm{deepseek did not do this, maybe worth mentioning ?}\askiad{In related work} 
In traditional expert parallelism, all experts are allocated the same number of replicas.
As explained in \S\ref{sec:background_moe}, this uniform allocation forces popular experts to drop excess tokens once their capacity is exceeded, slowing down convergence.

\NAME, by contrast, enables experts' replication to quickly follow their dynamic popularity.
Having eliminated data movement overheads during expert rebalancing, \NAME can update the expert placement on \textit{every iteration}.
\NAME reliably predicts expert popularity by mimicking the previous iteration's demand, and proportionally adjusts replication degrees.
This allows popular experts to scale their effective capacity without a latency penalty, dramatically reducing token drops and accelerating convergence.
% Our strategy is far simpler than coarser-grain predictive/heuristic-based replication schemes, yet much more effective.
% It is only after eliminating data movement overheads that we are allowed to adjust to fine-grained constraint-free placements.
\NAME's fine-grained, constraint-free strategy is far simpler, yet much more effective than coarser-grain predictive/heuristic-based schemes.

% \christos{maybe say that thsi is far simpler than any predictive or coarser-grain scheme and much more effective}

% \christos{this needs a little update. I assume the prev section (background) will have motivated the need to have non uniform and dynamic expert replication. There is related work on this after all. So what I would make here a big deal about is that you can you very fine-grain (quick) adaptivity! Once the overheads are down because of what you said above, you can adjust upon every single iteration. That should get you the best possible benefits for model accuracy. } \christos{for someone to buy that the per-iteration adjustment is a good idea, you need the text to refer back to figure 3 and say that trackign the prev iterations demand is the best way to approximate this curve. If you have some stats, throw them here}

% \christos{I'd also switch "system efficient adaptive replication" to no overhead adaptive replication (or low overhead) and "model efficient adaptive replication" to "continues adaptive replication" or "per-iteration adaptive replication" and then say that this will be the best for accuracy}
% \askiad{Agree. I wanted to differentiate with the titles in 3.3, 3.4. Ideas?}

\subsection{\NAME Design Overview}
\label{sec:design_overview}

% !TEX root =paper.tex
\begin{figure}[t]
  \centering
  \includegraphics[width=\linewidth]{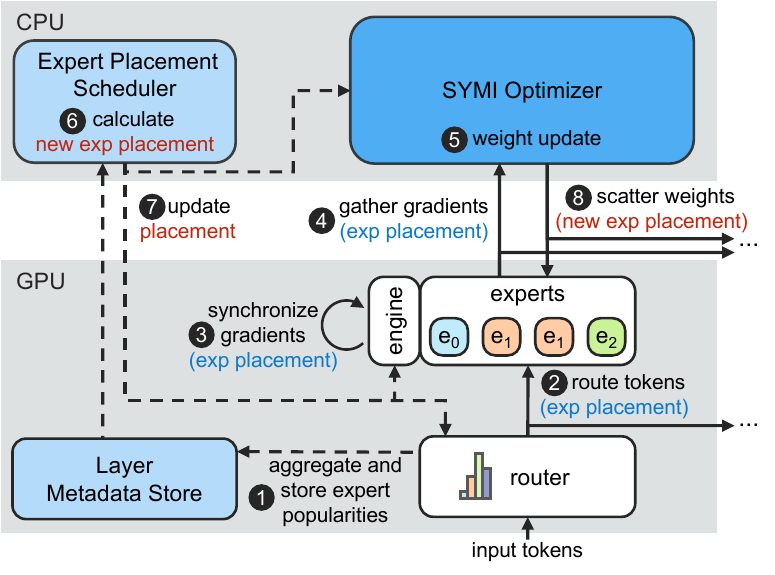}
  \caption{\NAME design block diagram. Showing data (solid lines) and metadata (dashed lines) flow in a single rank through a training iteration.}
  \label{fig:design_diagram}
\end{figure}

%\christos{I think you should say fig 5 shows the block diagram for Symi and enumerates the step to execute an entire training iteration and rebalancing step}
Figure~\ref{fig:design_diagram} shows \NAME's block diagram and enumerates the steps executed throughout an entire training iteration. 
We show a single MoE layer in a single rank (i.e., GPU)\footnote{Our design operates at the level of full expert instances, and techniques that split instances across ranks (e.g., tensor parallelism) are orthogonal to our approach, as discussed in Section~\ref{sec:discussion}. For simplicity of presentation here, we assume that each rank contains whole experts.}, and the process is repeated for all other layers and ranks. 
% \myzhao{I'd add a forward pointer to more discussion about this. It's not readily obvious why expert-sharded parallelism is orthogonal.}
The attention and normalization components are omitted; \NAME focuses only on the expert MLPs and does not change the rest of the Transformer layer as it is handled by current systems like DeepSpeed~\cite{pmlr22:rajbhandari_deepspeed-moe}.
% \christos{is there a baseline design here? if yes, you can say "does not change the rest of the transformer layer as it is handled by basline frameworks like DeepSpeed}
Specifically, \NAME introduces a set of core components that include the \textit{\NAME Optimizer}, which manages communication and expert updates, the \textit{Expert Placement Scheduler}, responsible for dynamic expert assignment, and a \textit{Layer Metadata Store} that tracks expert popularity. 
\NAME also extends the existing expert \textit{router} and runtime \textit{engine} to support dynamic expert replication.
% \askiad{Do we prefer capitalized names for each component?}\christos{either is fine. But keep it capitalized so people can tell that you are talking about the component}

Figure~\ref{fig:design_diagram} illustrates how these components operate over a full training iteration.
During the \textit{forward pass}, \NAME gathers expert popularity statistics and distributes tokens to the current (non-uniform) expert placement. % non-uniformly placed experts.
During the \textit{backward pass}, \NAME synchronizes expert gradients across each expert class's replicas. %according to their dynamic, non-uniform placement.
Finally, \NAME's \textit{optimizer step} gathers expert gradients, calculates the next iteration's expert placement, and performs rebalancing by distributing the updated weights according to the new placement.
% and allocates them according to the next iteration's expert placement to perform rebalancing.
%computes the updated weights, and allocates them back according to the next iteration's placement.

\noindent\textbf{Forward pass:} %\NAME gathers expert popularity information and distributes tokens to the non-uniformly placed experts.
% Forward
% 1. tokens from input/prev layer arrive
% 2. route tokens to classes, don't care about instance placement. distribute to replicas according to placement -- effectively multiplying capacity
% 3. collect counts, all-reduce, and store locally. every expert now has that info
Within each layer, the router takes in as input the batch token embeddings from the local attention block.
The router assigns tokens to expert classes, independently of their replication.
\numcircledtikz{1}~We extend the router to aggregate the number of tokens assigned to each expert class, via an all-reduce collective across all ranks.
The tensors participating in the all-reduce are small, containing only a single element for each expert class, so the overhead of the collective is negligible as shown in \S\ref{sec:eval_latency}.
% \myzhao{Is this an additional all-reduce that we introduce? If so, I'd suggest making this clear and adding a sentence on why this is cheap / easily hidden.} \shm{adding to Mark's comment. Is the AR just to get the counts ?}
\NAME stores the resulting globally-consistent expert popularity in the Layer Metadata Store, to be later used by the Expert Placement Scheduler.
\numcircledtikz{2}~The router maps tokens to expert classes, as usual.
\NAME then load-balances the tokens for a given expert class across its replicated instances.
Because \NAME replicates experts instances (i.e., the total capacity of each expert class) proportional to demand, \NAME minimizes drops and improves convergence.

%\NAME extends this assignment to expert instances, using the current replication degree of each expert.
%\myzhao{Not sure what "extends this assignment" means. "\NAME then load-balances the tokens for a given expert class across its replicated instances"?}
% \christos{maybe thank the fast adaptation frequency as well} 
%Thanks to the rapidly-adapting expert replication aligned to expert popularity,
% \christos{based on fine-grain expert popularity information?}, 
%\NAME effectively increases the capacity of each expert class in proportion to its demand, reducing drops and improving convergence.
% \myzhao{Highlight the impact of this on convergence / token drops. This is the key compared to static expert parallelism schemes.}
% \askiad{In fact, 3 and 7 happen before 2. But it messes up the flow of the text. And it doesn't matter what happens first anyway. Wdyt?} \christos{you can have a note about this at the end after you present the whole flow}

\noindent\textbf{Backward pass:} %\NAME synchronizes expert gradients according to their dynamic, non-uniform placement.
% Backward
% 4. sync gradients across experts of same class. secXX explains later how this is done efficiently
\numcircledtikz{3}~During the backward pass, the runtime engine (e.g., DeepSpeed) performs an all-reduce step to synchronize expert gradients.
While this all-reduce exists in current engines, \NAME changes the ranks that contain instances for each expert class on a per-iteration basis.
Thus, \NAME introduces a new locality-enhanced all-reduce implementation (Section~\ref{sec:comm}) 
to efficiently synchronize gradients across the dynamic communication groups.
% that dynamically adapts communication groups to synchronize across the appropriate ranks.

%While this is an existing collective in baseline systems, the participating ranks for each expert may change per iteration.
%In \NAME, the engine consults the current expert placement to use the appropriate distributed communication groups, consisting of ranks that host instances of the same expert class.
%We analyze our low-overhead all-reduce and efficient communication group manipulation in Section~\ref{sec:comm}.
% \myzhao{Again, not clear if this is a new collective that we introduce/require, or if this is already existing.}

\noindent\textbf{Optimizer step:} %The \NAME Optimizer gathers expert gradients, computes the updated weights, and allocates them back according to the next iteration's placement.
% Opt 
% 5. gather gradients from experts (experts follow current iteration placement)
% 6. compute the updated weights in the CPU
% 7. calculate new placement. practically, this step has ran already, not in critical path
% 8. update placement to different components (router for routing, engine for backdrop, optimizer)
% 9. scatter expert model weights to GPUs using the new distribution.
\NAME's Optimizer manages the decoupled optimizer state for all expert instances.
The \NAME Optimizer is offloaded to host memory and is uniformly partitioned for all experts across all nodes.
% \new{
We prove this strategy is latency-optimal in Appendix~\ref{apx:opt_part}).
% }
% \rem{(in Appendix \ref{apx:opt_part} we prove this is the optimal strategy)}
% \christos{either here or somewhere in 4.1 you need to say why distributing across all nodes is a good idea. Why not half?}
% \myzhao{I'd add a high-level insight (half a sentence) right before the fwd ptr}
\numcircledtikz{4}~The \NAME Optimizer in each node gathers its corresponding gradient partitions,
\numcircledtikz{5}~and uses them to perform the optimizer step and produce the updated weights (as in baseline systems).
\numcircledtikz{6}~Meanwhile, the Expert Placement Scheduler (\S\ref{sec:design_sched}) collects this iteration's expert distribution from the Layer Metadata Store and calculates the expert instance allocation for the next iteration\footnote{In practice, step 6 may execute earlier, even right after step 1.}.
\numcircledtikz{7}~\NAME updates the \NAME Optimizer, runtime engine, and router with the new expert placement.
\numcircledtikz{8}~The \NAME Optimizer finally sends the updated expert weights to slots according to the new, rebalanced schedule.
%t this point, the expert rebalancing has been completed, without requiring any extra data movement.

% While steps \numcircledtikz{4} and \numcircledtikz{8} also occur in baseline systems, \NAME changes the ranks participating in each collective.  %locality of the collectives might change.
We next prove that our collective communication requires no additional data movement and has equivalent communication cost
% with static and uniform expert replication. 
compared to the equivalent steps \numcircledtikz{3}, \numcircledtikz{4} and \numcircledtikz{8} of baseline static systems.

\subsection{No-overhead Adaptive Replication}
\label{sec:design_comm}

% !TEX root =paper.tex
\begin{figure}[t]
  \centering
  \includegraphics[width=\linewidth]{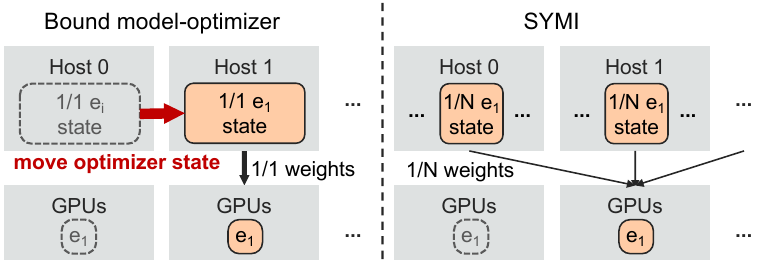}
  \caption{Current systems (left) bind optimizer state to expert instances, requiring costly optimizer state migration during expert rebalancing. \NAME (right) keeps the optimizer static and eliminates this overhead.}
  \label{fig:design_rebalance}
\end{figure}

% Total data movement.
Current systems co-locate model and optimizer (partitioned or not) state.
The optimizer state accounts for the majority of an expert's memory footprint (\S\ref{sec:background_adaptive}: $8\times$ more than model weights). 
% \myzhao{A better term than "traditional designs"? Current systems?}
Consequently, current adaptive expert replication solutions suffer from significant communication overhead moving optimizer state together with its respective expert's weights.
This overhead becomes especially problematic given the need to continuously rebalance expert instances. %, as discussed in Section~\ref{sec:background_adaptive}.
%highly dynamic expert distribution during training (\S\ref{sec:background_moe}~Figure~\ref{fig:motivation_distr}).

As shown in Figure~\ref{fig:design_rebalance}, \NAME avoids this overhead entirely.
% by decoupling optimizer and parameter states.
\NAME develops a clean separation between \textit{static optimizer state in host memory} and \textit{dynamic expert weights in accelerators}.
% \NAME \textit{statically manages optimizer state in host memory} and \textit{dynamically adapts expert weights across accelerators}.
Similar to ZeRO-1~\cite{rajbhandari2020zero}, \NAME offloads optimizer state to host memory.
However, unlike ZeRO, \NAME decouples the offloaded optimizer from expert instance placement;
\NAME uniformly partitions each expert's optimizer across all $N$ nodes, and changes the expert class assigned to each GPU expert slot on every iteration.

% The optimizer placement is static, regardless of expert placement, allowing \NAME to \textit{remove the optimizer migration overhead}.

Our design successfully eliminates the optimizer migration overhead.
% However, in contrast to current systems, \NAME can change the expert class assigned to each GPU expert slot, \textit{every iteration}.
The remaining challenge is to avoid introducing overhead when relocating experts.
\NAME's insight is to \textit{perform rebalancing by leveraging existing data movement}, rather than introducing additional communication to shuffle weights across GPUs.
In particular, the \NAME Optimizer disregards the previous placement after the optimizer step, and materializes the new expert placement by transferring the updated weights to each expert slot according to the next iteration's rebalanced schedule.

In total, the communication needed to rebalance expert state can be split in two phases.
The \textit{Grad Communication Phase} synchronizes and collects expert gradients from expert instances to the \NAME Optimizer, based on the previous expert placement.
The \textit{Weight Communication Phase} distributes the updated expert weights from the \NAME Optimizer to expert slots based on the new expert placement.

% \christos{I think this section could use a before and after block diagram. The before can be dynamic expert replication where you move both experts and optimizer state. If these diagrams are small, you can also have two befores (one static assignment). The edges in the diagrams can be annoated to indicate volumes of data}

% \askiad{Thoughts on moving the math to the appendix and presenting here the results? Maybe do it only for (III) which is the heaviest?} \shm{I feel this is the key to proving that the cost is minimal}
To show that \NAME introduces no overheads, we compute (I) the total optimizer state footprint, (II) the total data IO in each phase, and (III) the communication cost of each phase using \NAME and a baseline design with static expert replication and offloaded optimizer like DeepSpeed~\cite{ics23:singh_deepspeed-ted}.
In the baseline static design, each expert is replicated a constant number of times, and the corresponding optimizer state is uniformly sharded across the nodes hosting that expert.

\begin{table}[h]
    \centering
    \begin{tabular}{ c | l }
        % \hline
        % Notation & \\
        % \hline \hline
        $N$ & \# nodes in the training cluster \\
        $E$ & \# expert classes \\
        $s$ & \# expert slots per rank \\
        $r$ & \# expert replicas (static baseline) \\
        $r_i$ & \# expert replicas for expert $e_i$ (\NAME) \\
        $BW_{pci}$ & local GPU-CPU interconnect (e.g., PCIe) bandwidth \\
        $BW_{net}$ & cross-node GPU-GPU network (e.g., IB) bandwidth \\
        $G$ & gradients data size for one expert instance \\
        $W$ & weights data size for one expert instance \\
        $O$ & optimizer state data size for one expert class \\
        $M$ & total optimizer memory footprint \\
        $D_{G/W}$ & total data transferred in each phase \\
        $T_{G/W}$ & communication cost per rank in each phase \\  
        % \hline
    \end{tabular}
    \caption{Variable Definitions}
    \label{tab:definitions}
\end{table}

We use the  notation\footnote{For simplicity, this model assumes that each node contains a single GPU rank with $s$ expert slots. This translates to $s$ total slots across a node's NVLink-connected GPUs, with experts possibly sharded via tensor parallelism.} in Table \ref{tab:definitions}. Notice that the total number of expert instances in the system are:
% $rE = sN$ for the static baseline, and $\sum_{e_i}r_i = sN$ for \NAME.
\begin{align*}
 rE &= sN,~ \text{for the static baseline} \tag{1} \\
 \sum_{e_i}r_i &= sN,~ \text{for \NAME} \tag{2} \\
\end{align*}

To illustrate scale, we often accompany the equations with a representative example extending a GPT3-175B model ($G=W=3.375$GB and $O=27$GB)~\cite{brown2020language} with $E=64$ experts~\cite{liu2024deepseek}, trained in a cluster with $N = 2048,~s=2,~BW_{pci}=64\text{GB/s},~BW_{net}=400\text{Gbps}$~\cite{nvidiaH100}. 
% \christos{it is better if you say that the example numbers are for a model people knwo and a system people know}

\paragraph{(I) Optimizer Memory Footprint}

We compare the total memory footprint of the \NAME Optimizer to the static baseline. The static baseline partitions the optimizer of each expert $r$-ways, following expert data parallelism. \NAME partitions the optimizer of each expert across all $N$ nodes. 

Both designs have the same memory footprint:
\begin{align*}
M^{static}
=&
E \frac{1}{r} r O = EO \\
M^{\NAME}
=&
E \frac{1}{N} N O = EO \\
\end{align*}
which is ${\sim}1.7$TB per layer, equally distributed over the host memory of the cluster.

\paragraph{(II) Data Transferred} We compute the total data involved in communication, regardless of the collective implementation (e.g. different all-reduce algorithms).

For the \textit{Grad Communication Phase}, for all experts each optimizer partition needs as input the corresponding gradient shard, synchronized across all expert replicas: for each expert ($e_i$), all expert replicas ($r_i$) participate in exchanging their gradients shards ($\frac{G}{\# partitions}$) corresponding to each optimizer partition ($\# partitions$).

{\par\sloppy 
For the \textit{Weight Communication Phase}, each expert instance needs to receive the concatenation of all weight shards: each expert ($e_i$), receives the full partitioned weights
($\# partitions~\times~\frac{G}{\# partitions}$) in each replica ($r_i$).
\par}

The above are formulated as:
\begin{align*}
D_G^{static}
=&
E r \frac{G}{r} r = rEG \stackrel{(1)}{\varequals{5pt}} sNG \\
D_W^{static}
=&
E r \frac{W}{r} r = rEW \stackrel{(1)}{\varequals{5pt}} sNW \\
D_G^{\NAME}
=&
\sum_{e_i} r_i \frac{G}{N} N \stackrel{(2)}{\varequals{5pt}} sNG \\
D_W^{\NAME}
=&
\sum_{e_i} N \frac{W}{N} r_i \stackrel{(2)}{\varequals{5pt}} sNW \\
\end{align*}

showing that an equal volume of data (27TB total, involving all nodes and all networks) is moved on every iteration in both \NAME and the static baseline.

\paragraph{(III) Communication Cost}

At this point, we have proved that \NAME communicates the same volume of data as static expert replication.
However, the locality of the optimizer state and expert slots differ in our design.
We next show that this introduces negligible communication overhead.
%At the same time, \NAME (unlike static design) does not sacrifice samples-to-convergence while current approaches that right-size expert replicas incur immense overhead.

% \myzhao{This is the key, and should be more clear. 
% We ensure that the volume of data communicated across GPUs is the same. 
% However, the locality of the optimizer state and expert slots are different. We show that this introduces negligible communication overheads.}
% \myzhao{Actually, I'd maybe frame this differently  with 2 baselines. The naive baseline has minimal network requirements, but significantly sacrifices samples-to-convergence. Current works which right-size expert popularity can maximize samples-to-convergence, but take a *lot* of network overheads. Here, we can also maximize samples-to-convergence, but require much less network costs. Experimentally, we show that all of this leads to better time-to-convergence. Maybe frame the motivation sec like this?}

In the \textit{Grad Communication Phase}, each rank collects the synchronized (reduced) expert gradient shards corresponding to the local optimizer partitions via the backend network, and then transfers these gradient shards to host memory via the CPU-GPU interconnect (e.g., PCIe).

Similarly, in the \textit{Weight Communication Phase}, the expert optimizer lands the updated weight shards in GPU HBM via PCIe, and then transfers the weight shards to any corresponding remote expert replicas through the backend network.

The full computations are provided in Appendix~\ref{apx:comm}, and the resulting expressions are summarized below:

\begin{align*}
% T G static
T_{G}^{static}
&=~
\frac{E}{N} \frac{G}{BW_{pci}}
+
\frac{sN-E}{N} \frac{G}{BW_{net}} \\
% T W static
T_{W}^{static}
&=~
\frac{E}{N} \frac{W}{BW_{pci}}
+
\frac{sN-E}{N} \frac{W}{BW_{net}} \\
% T G Symi
T_{G}^{\NAME} 
&=~
\frac{E}{N} \frac{G}{BW_{pci}}
+
\frac{sN-s}{N} \frac{G}{BW_{net}} \\
% T W Symi
T_{W}^{\NAME}
&=~
\frac{E}{N} \frac{W}{BW_{pci}}
+
\frac{sN-s}{N} \frac{W}{BW_{net}} \\
\end{align*}

We compare the resulting formulas and find that \NAME presents only marginally more communication cost (due to the reduced expert-optimizer locality).
This small increase is equal to $\frac{\Delta T}{T^{static}} = \frac{E-s}{sN - E(1 - \frac{BW_{net}}{BW_{pci}})}$.

In practice, this extra communication overhead between the static baseline and \NAME is negligible. In our example, \NAME would incur only $1.52\%$ extra communication cost per iteration (${\sim}0.273$s vs ${\sim}0.269$s total communication).
% Notably, the proportional discrepancy in iteration latency is even lower as \NAME does not alter the main computation parts of each iteration, and all added components have negligible overhead (\S\ref{sec:eval_latency}). 
% \myzhao{not sure I follow, why would this make it even lower?}
In summary, \NAME enables experts to be efficiently rebalanced at a per-iteration granularity.

% In contrast to approaches that migrate optimizer state alongside expert weights—incurring up to $\frac{E \cdot O}{BW_{\text{net}}} = 34.56,\text{s}$ of additional rebalancing overhead—\NAME avoids rebalancing cost entirely.

% \myzhao{in terms of latency per iteration, at the benefit of much fewer dropped tokens (and faster convergence)}
% and 256/1024/4096 nodes would incur only $5.94\%$/$1.47\%$/$0.37\%$ extra communication cost, respectively.
% \askiad{How's the above argument? We need no flags raised here. The extra communication is honestly negligible. It is $\frac{delta}{old comm} = \frac{E-s}{sN - E(1 - \frac{BW_{net}}{BW_{pci}})}$ and decreases almost linearly with $N$. Note that BW doesn't matter if the PCI/net BW ratio is constant (it is across generations) -- is this worth mentioning?}

% \christos{I'd move the long version of the formulas to an appendx and keep here a short version with the intuition behind it! This is a very long set of formulas and many readers will space out. } \askiad{Done, only final results here. Or even shorter?} \christos{it depedns on how you are doing on space}

% \christos{you are also missing a statement at the very end that says you just made rebalancing low cost enough to do it on every iteration}

% \subsection{Continuous Adaptive Replication}
\subsection{Expert Placement Manipulation}
\label{sec:design_sched}

As discussed in Section~\ref{sec:background_moe}, current systems impose a fixed capacity limit on the number of tokens that can be routed to each expert class. 
Expert capacity aims to avoid long iterations and idle resources due to the high variance in expert popularity.
However, this causes dropped tokens, which harms training convergence. % dropping excess tokens damages convergence.

Expert capacity in current systems is defined as:
\begin{align*}
\mathrm{capacity}(e_i)
=&~\mathrm{capacity\_factor} \times \frac{\mathrm{tokens\_per\_batch}}{E} \\
\stackrel{(1)}{\varequals{5pt}}& \underbrace{\mathrm{capacity\_factor} \times \frac{\mathrm{tokens\_per\_batch}}{s N}}_{\mathrm{slot\_capacity}}  \times r \\
\end{align*}
% where $r$ is the constant replication degree of each expert class.
Setting the $\mathrm{capacity\_factor}$ to $1.0$ results in the lowest iteration latency, but also the highest drop rate.

Conversely, \NAME allows experts to be replicated non-uniformly, effectively scaling each expert class's total capacity by the dynamic number of its assigned replicas ($r_i$):
\begin{equation*}
% \boxed{
\mathrm{capacity^{\NAME}}(e_i) = \mathrm{slot\_capacity} \times r_i
% }
\end{equation*}

If each expert's replication degree is proportional to its popularity (barring rounding errors), tokens can be routed to their assigned expert classes with minimum drop rate and minimum iteration latency.
Popular expert classes have enough instances to serve all their assigned tokens without token drops or extra latency.
In this ideal scenario, the $\mathrm{capacity\_factor}$ is irrelevant.

However, the system can precisely capture each iteration's expert popularity only after the router assignment.
The cost of reshuffling experts between expert assignment and token routing would be prohibitive. 
Thus, \NAME relies only on past popularity information and materializes the expert placement in the previous iteration, avoiding this cost. 

Specifically, after the router assignment, \NAME invokes an all-reduce collective to aggregate the number of tokens assigned to each expert class.
The resulting array of sums represents the current iteration's expert popularity.
\NAME stores the popularity array into the local rank's Layer Metadata Store.
% The Layer Metadata Store is an abstraction, local to each rank, collecting necessary past popularity information.
The Expert Placement Scheduler retrieves its desired input information from the Layer Metadata Store, and produces the expert placement schedule for the next iteration.
The overhead of those added components is negligible (see \S\ref{sec:eval_latency}).
% \myzhao{See earlier comment on if this can be easily hidden. Also, may be worth factoring this into your previous calculations on addtl nw costs?}

The Expert Placement Scheduler can leverage any historical popularity information to derive the next iteration's placement.
In this paper, we chose the simple, yet very effective policy; \textit{expert placement mimics the previous iteration's popularity}.
As discussed in \S\ref{sec:background_moe}, expert popularity can shift dramatically within as few as 3 iterations, highlighting the need for per-iteration rebalancing. 
At the same time, expert popularity distributions are smooth enough so that the previous iteration serves as a reliable proxy for the next.
In our evaluation, we demonstrate that our policy is sufficient for replication to accurately match popularity, enabling \NAME to drop $43\%$-$64\%$ fewer tokens than systems that rebalance experts on 10-100 iteration intervals (\S\ref{sec:eval_convergence}). 
% We also discuss different potential replication policies in \S\ref{sec:discussion}.

The Expert Placement Scheduler computes how expert instances are distributed across devices in each iteration.
The scheduler assigns replicas to experts in proportion to their captured popularity, assigning at least one instance per expert so that all experts remain reachable.
The scheduler then places the assigned instances \textit{contiguously across slots}, favoring placement of same-class experts within the same rank.
The full algorithm is provided in Appendix~\ref{apx:algo_scheduler}.
The Expert Placement Scheduler runs locally on every rank.
Because the Expert Placement Scheduler's algorithm is deterministic, the only control-flow coordination needed between ranks is to aggregate the input expert popularities.
% In more complex potential scheduling algorithms that would use different input data, input synchronization can even be performed asynchronously, off the critical path of each training iteration.

\section{\NAME Collective Communication}
\label{sec:comm}

We implemented \NAME on top of DeepSpeed~\cite{rasley2020deepspeed}.
% Unfortunately, DeepSpeed (and other current systems) do not have the collective communication mechanisms needed to support \NAME's dynamically-changing expert data parallel groups.
This section presents \NAME's collective communication mechanisms that enable the per-iteration adaptive replication described in Section~\ref{sec:design}.
We show how \NAME synchronizes gradients across instances of the same expert class (\S\ref{sec:comm_allreduce}), while efficiently managing distributed communication groups (\S\ref{sec:comm_group}).
We then present how \NAME collects the gradients (\S\ref{sec:comm_gradgath}) and distributes the updated weights to expert slots (\S\ref{sec:comm_scatweights}).

\subsection{Intra+Inter Rank All-Reduce}
\label{sec:comm_allreduce}
% !TEX root =paper.tex
\begin{figure}[t]
  \centering
  \includegraphics[width=\linewidth]{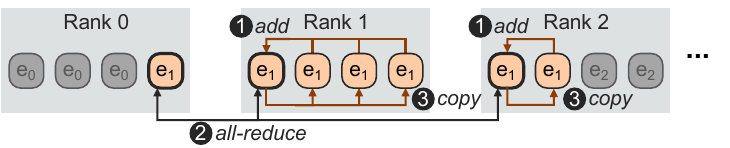}
  \caption{\NAME extends the all-reduce implementation allowing intra-rank expert replication.}
  \label{fig:impl_allreduce}
\end{figure}

% Multiple expert slots per rank. Explain how we implement collectives using intra-rank communication (local aggregation -> cross rank collective -> local copy).
% The \NAME Optimizer uses the \textit{Grad Communication Phase} to synchronize and collect the expert gradients.
% \shm{did not grok the next paragraph. If we say reduce scatter, you should need a bcast at the end which will be a multiply by r, if doing all-reduce you get what you have ... but the value in time of each operation is not the same} 
In principle, a partitioned optimizer synchronizes gradients with communication size of $\frac{(r-1)G}{r}$ % CANTRIM
% ; each of $r$ data-parallel ranks needs to receive the $\frac{G}{r}$ gradients from the remaining $r-1$ ranks 
(as in ZeRO-2~\cite{atc21:ren_zero-offload}).
This is a reduce-scatter collective -- the first half of an all-reduce.
In practice, existing frameworks perform the full all-reduce collective across the $r$ data-parallel partitions with communication size of $2\frac{(r-1)G}{r}$.
% A part of this all-reduce cost is often hidden, as gradient communication is overlapped with computation.

% \myzhao{This paragraph has a lot of extraneous information. Can probably collapse to a single sentence that says something along the lines of "To partition optimizer state, each iteration needs to perform an all-reduce across expert instances"}

Existing all-reduce collective implementations synchronize tensors across different ranks, but not within them~\cite{paszke2019pytorch, nccl},
preventing experts from being freely allocated to expert slots.
Each expert can only be replicated across different ranks up to $N$ times, instead of up to $sN$ allowing multiple instances of the same expert class on the same device.
This constraint makes expert scheduling complicated and often leads to sub-optimal schedules.
% \footnote{For example, in a system with 8 GPUs, 2 slots per GPU and 4 experts, an expert placement schedule of 5, 5, 4, 2 instances is impossible to materialize.}.
We have empirically found that this constraint can increase token drops by up to $20\%$.

\NAME proposes a novel all-reduce implementation that allows simultaneously both inter- and intra- rank expert data parallelism, removing replication restrictions. %it lifts any replication restrictions.
% and in many cases leads to considerable performance benefits.
% \shm{ the referenced figure is a little confusing. Can we lable the boxes [eN, SN] since the reduction is really across slots, not necessarily experts in the replicated case}
Figure~\ref{fig:impl_allreduce} illustrates how our proposal manages experts replicated both within and across ranks.
\numcircledtikz{1}~For a given expert class, each rank elects a slot representative and the remaining expert slots within the rank add their tensors to the representative.
\numcircledtikz{2}~Then, an inter-rank all-reduce is applied only across each rank's representative slots.
\numcircledtikz{3}~Finally, the representative slot in each rank normalizes and copies the all-reduced gradients to the remaining slots, completing the all-reduce.

Besides enabling arbitrary expert placement schedules, our all-reduce implementation
% leverages the fast HBM bandwidth for the add and copy operations, and
synchronizes instances of each expert class with less inter-node network traffic
compared to schedules that would have to spread equally-as-many instances across different ranks.
\NAME leverages this property in full; the Expert Placement Scheduler (\S\ref{sec:design_sched}) assigns expert instances contiguously first across slots within a rank and then across ranks.
% across the slot-rank domain, minimizing the spread of each expert. \myzhao{lots of jargon (slot-rank domain, spread). not sure what this means. can you simplify? Why is minimizing "spread" helpful?}
% The inta+inter rank all-reduce implementation also extends to dense models and allows system designers to fully utilize HBM capacity with data parallelism before crossing the boundaries of a rank. \myzhao{same, not sure what this means. drop?}
% \myzhao{this is pretty hard to understand and could use a pass. At least use the figure better with numbers/labels.}

\subsection{Communication Group Management}
\label{sec:comm_group}

As replication patterns evolve, the required all-reduce communication in the backward pass may involve a varying set of ranks for each given expert. 
NCCL mandates that collective operations occur over all ranks in explicitly defined communication groups~\cite{torchgroups}.
Consequently, dynamic replication requires creating a new communication group for each expert, at each layer, potentially on every iteration.
This operation includes blocking, single-threaded synchronization and may take more than 1,000 seconds in large clusters (e.g., $N=2048$)~\cite{jiang2024megascale}.
To address this prohibitive overhead, \NAME pre-registers all necessary communication groups at initialization time.
To avoid registering all possible $2^N$ rank combinations, we leverage that the Expert Placement Scheduler (\S\ref{sec:design_sched}) assigns experts contiguously across ranks.
Thus, we only register groups of consecutive ranks, requiring \NAME to only manage $\frac{N(N-1)}{2}$ groups.
This allows us to reuse the same pre-initialized groups across different experts and layers, ensuring zero group-creation overhead during training.

%Because Algorithm~\ref{alg:scheduler} assigns experts contiguously across ranks, we only need to register groups for contiguous ranks.
%As a result, this allows \NAME to manage $\frac{N(N-1)}{2}$ groups, rather than the intractable set of all possible $2^N$ rank combinations.
%This allows us to reuse the same pre-initialized groups across different experts and layers, ensuring zero group-creation overhead during training. %\christos{as a reader I would like to know what are reasonably ranges for N so taht I know how big these numbers are. given some range}
 
\subsection{Gradient Communication Load-Balance}
\label{sec:comm_gradgath}
% Once gradient synchronization is complete, the \NAME Optimizer on each rank proceeds to update its state by fetching its respective partition of the gradients.
% Algorithm~\ref{alg:gathergrad} shows how the \NAME Optimizer efficiently selects a unique expert instance to collect each expert class's gradient shard.
% The \texttt{\small get\_source} function assigns a single source rank to a given expert and optimizer destination rank.
% To minimize the amount of inter-rank communication, we prioritize local communication if the expert is local to the optimizer.
% Otherwise, \texttt{\small get\_source} round-robins across different expert instances to avoid communication hotspots and bottlenecks.
%%%
Once gradient synchronization is complete, the \NAME Optimizer on each rank proceeds to update its state by fetching its corresponding gradient partitions. 
To coordinate this transfer efficiently, \NAME selects a unique expert instance as the source for each expert class’s gradient shard.
To minimize communication cost, \NAME prioritizes local expert–optimizer transfers whenever possible, whereas remote transfers are distributed in a round-robin fashion across available expert replicas to avoid network contention and hotspots.
The full gradient collection algorithm is provided in Appendix~\ref{apx:algo_gradgath}.

\subsection{New Expert Placement Materialization}
\label{sec:comm_scatweights}
Finally, the \NAME Optimizer computes the updated weights and distributes them to expert slots according to the next iteration's expert placement.
\NAME implements both the gradient collection (\S\ref{sec:comm_gradgath}) and the weight update with batch point-to-point communication~\cite{torchp2p}.
\NAME identifies the participating send/receive ranks and issues a \texttt{\small batch\_isend\_irecv} operation across all experts.
% we use batched p2p (isend/irecv) - p2p skipped if local

% \subsection{Expert Placement Scheduler}

% \input{sections/implementation}
\section{Evaluation}
\label{sec:evaluation}
\noindent\textbf{Experimental Setup.}
We compare \NAME to two state-of-the-art baselines.
DeepSpeed~\cite{rasley2020deepspeed} represents the state-of-the-art \textit{static} baseline, which does not perform any adaptive replication.
We also compare against the state-of-the-art \textit{adaptive replication} baseline, FlexMoE~\cite{sigmod23:nie_flexmoe}.
We ran experiments on an Azure cluster with 16 NC24ads-v4 instances.
Each instance contains an NVIDIA A100 80GB GPU with a 32 GB/s PCIe 4.0 interconnect and a 100Gbps ConnectX-5 NIC. 

All systems set the capacity\_factor to 1.0, the auxiliary loss coefficient to $10^{-5}$, and use Top-$k$=1 routing with 16 expert classes and 4 expert slots per GPU, totaling 64 expert instances for every layer.
DeepSpeed allocates an equal number of expert instances (4) for each expert class, distributed across different ranks as DeepSpeed does not support intra-rank expert data parallelism.
For all systems, we set the tensor and pipeline parallelism degrees to 1, as these strategies are orthogonal to our design.
For non-expert components, we use a data parallelism degree of 16 to match the world size~\cite{sc21:narayanan_training-megatron}.
We configured DeepSpeed with ZeRO-1 (i.e., optimizer offloaded to CPU DRAM 
and evenly sharded across the nodes containing instances of the same expert).
\NAME partitions the optimizer of all experts uniformly across all ranks.

Since FlexMoE does not have an open source implementation, we implemented FlexMoE's expert scheduling policy over \NAME. 
% \new{
As in the original implementation, our FlexMoE implementation ties each expert's full optimizer state to its expert instances. 
% Gradients are reduced before the optimizer update (similar to DeepSpeed/\NAME).
Since the optimizer must remain local to its expert, during rebalancing the entire optimizer state is transferred to nodes that did not previously host that expert.
% FlexMoE ties the optimizer state to each expert. Since the optimizer is local, rebalancing requires transferring the full optimizer state to any node that gains a new expert.
% }
% \rem{As with the original implementation, our FlexMoE implementation couples the optimizer state to each expert instance, and each expert rebalancing requires transferring the entire optimizer state.}
To explore the accuracy and latency tradeoff between different FlexMoE rebalancing strategies, we followed the rebalancing frequencies reported in the original paper~\cite{sigmod23:nie_flexmoe} and triggered rebalancing every $i=$ 10, 50, or 100 iterations.

We used a standard GPT architecture~\cite{brown2020language}, with three base-model sizes; GPT- Small, Medium, and Large, with 125M, 350M, and 760M parameters, respectively.
We are training on the MMLU dataset~\cite{Hendrycks2020-mmlu}.
We set sequence length to 512 and global batch size to 64.

% \christos{this will likely be trimmed} 
% We first evaluated how \NAME's per-iteration adaptive replication can enable overall faster \textit{time-to-convergence} compared to baseline systems (\S\ref{sec:eval_timetoconv}).
% To better understand why this is the case, we next analyze the two factors that contribute to time-to-convergence.
% First, we explore how \NAME's ability to adapt expert placement \textit{each} iteration minimizes the number of tokens dropped each training iteration, improving the convergence rate on a per-iteration basis (\S\ref{sec:eval_convergence}).
% Secondly, we explore how \NAME does not introduce latency overheads, even compared to DeepSpeed's static replication (\S\ref{sec:eval_latency}), validating the theoretical analysis in \S\ref{sec:design_comm}.

\subsection{Time to Convergence}
\label{sec:eval_timetoconv}

% !TEX root =paper.tex
\begin{table}[t]
    \centering
    \caption{Total training time (in minutes) to reach target loss.}
    % \resizebox{.5\textwidth}{!}{
    \begin{tabular}{c c c c c}
    \toprule
    \thead{DeepSpeed} & \thead{FlexMoE-100} & \thead{FlexMoE-50} & \thead{FlexMoE-10} & \thead{\textbf{\NAME}} \\
    % DeepSpeed & FlexMoE-100 & FlexMoE-50 & FlexMoE-10 & \textbf{\NAME} \\
    \midrule
    \thead{$147.84$} & \thead{$145.42$} & \thead{$141.60$} & \thead{$138.61$} & \thead{\textbf{102.68}} \\
    \bottomrule
    \end{tabular}
    %}
\label{tab:eval_time}
\end{table}

Time-to-convergence represents the end-to-end metric that measures how fast we can train a model to some target loss. Time-to-convergence directly relates to the cost of training as well. 
Table~\ref{tab:eval_time} shows the time-to-convergence needed to reach a target loss value of 4.0, training a GPT-Small (125M) parameter model.
\NAME is able to improve the time-to-convergence by $30.5\%$ compared to DeepSpeed.
While FlexMoE with high rebalancing frequency is able to converge faster than DeepSpeed, \NAME is still faster by $25.9\%$ to $29.4\%$.
This result is a combination of \NAME's improved convergence performance (fewer iterations to target loss) and system performance (lower iteration latency).
Next, we separately analyze these two components.

\subsection{Convergence Evaluation}
\label{sec:eval_convergence}

% !TEX root =paper.tex
% \begin{figure}[t]
%   \centering
%   \includegraphics[width=.95\linewidth]{figures/img/convergence.pdf}
%   \caption{GPT-Small training loss for \NAME, DeepSpeed, and FlexMoE with different rebalancing intervals. \NAME achieves faster convergence than static replication systems like DeepSpeed and coarse-grained adaptive replication solutions.}
%   \label{fig:eval_convergence}
% \end{figure}

% \begin{figure}[t]
%   \centering
%   \includegraphics[width=.99\linewidth]{figures/img/survived_tokens_total.pdf}
%   \caption{Percentage of survived tokens across iterations. Static expert replication suffers from token drops, while infrequent rebalancing cannot adjust well to the dynamic expert activation distribution.}
%   \label{fig:eval_survived}
% \end{figure}

\begin{figure}[t]
  \centering
  \begin{minipage}{\linewidth}
    \centering
      \includegraphics[width=0.995\linewidth]{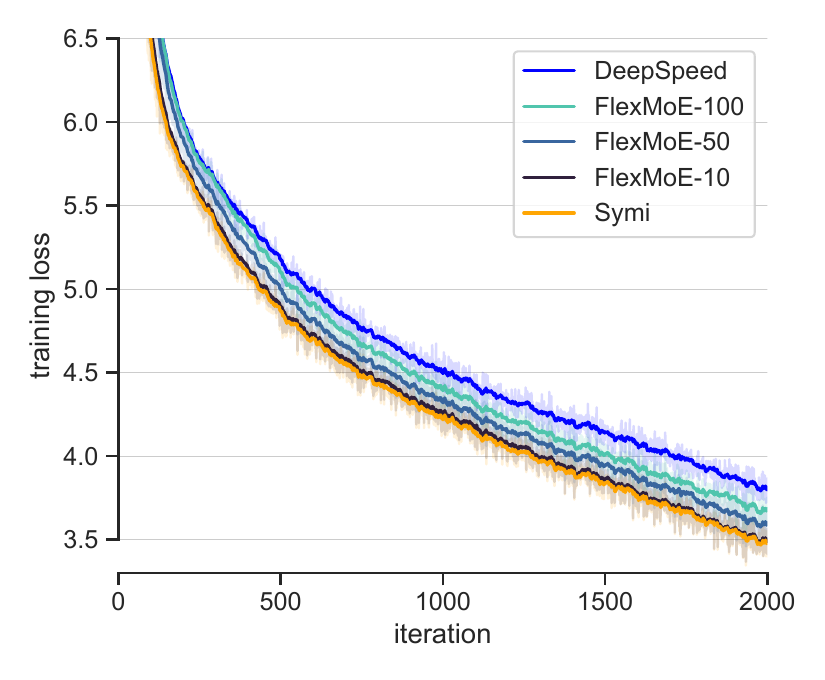}
      \vspace*{-18pt}      
      \caption{GPT-Small training loss for \NAME, DeepSpeed, and FlexMoE with different rebalancing intervals. \NAME achieves faster convergence than static replication systems like DeepSpeed and coarse-grained adaptive replication solutions.}
      \label{fig:eval_convergence}
  \end{minipage}
  
\vspace{12pt}

  \begin{minipage}{\linewidth}
    \centering
      \includegraphics[width=0.995\linewidth]{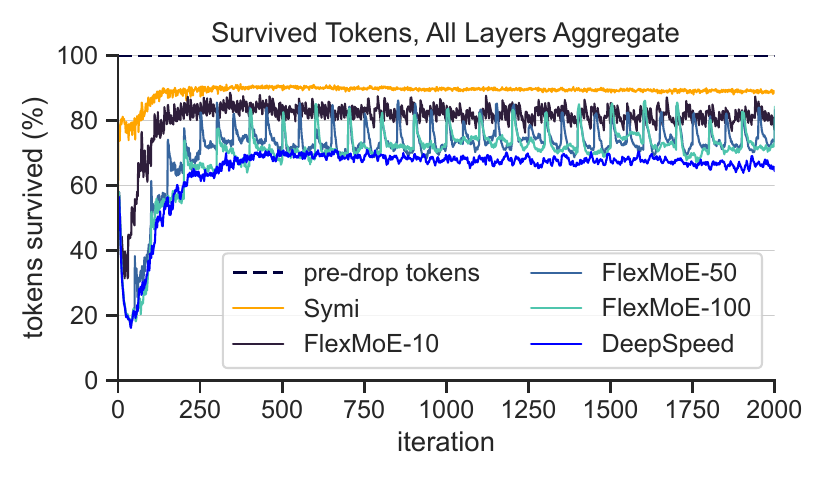}
      \vspace*{-18pt}
      \caption{Percentage of survived tokens across iterations. Static expert replication suffers from token drops, while infrequent rebalancing cannot adjust well to the dynamic expert activation distribution.}
      \label{fig:eval_survived}
  \end{minipage}
\end{figure}
% \input{figures/tex/survived}

% !TEX root =paper.tex
\begin{figure*}[t!]
  \centering
  \includegraphics[width=0.32\linewidth]{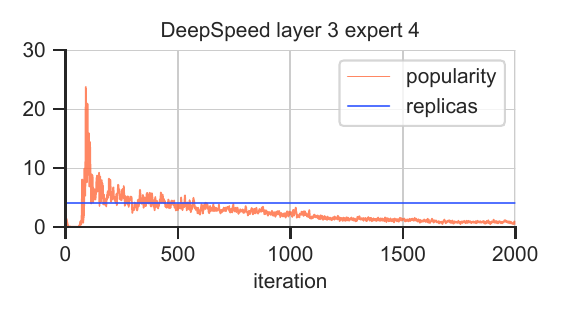}
  \includegraphics[width=0.32\linewidth]{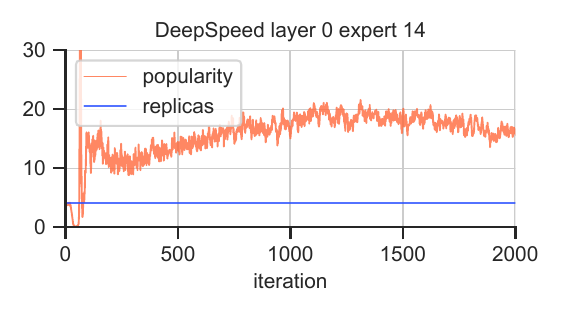}
  \includegraphics[width=0.32\linewidth]{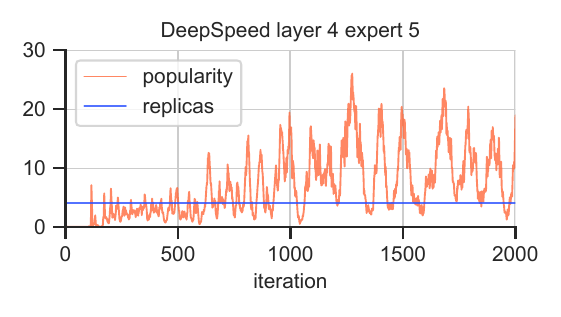}
  \includegraphics[width=0.32\linewidth]{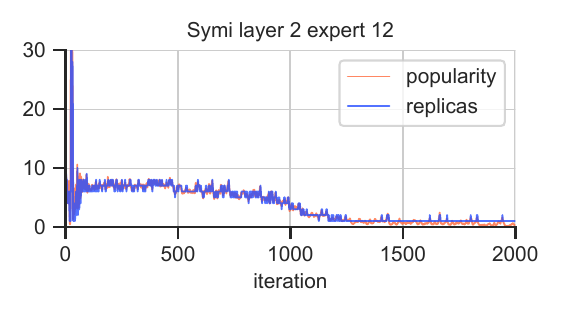}
  \includegraphics[width=0.32\linewidth]{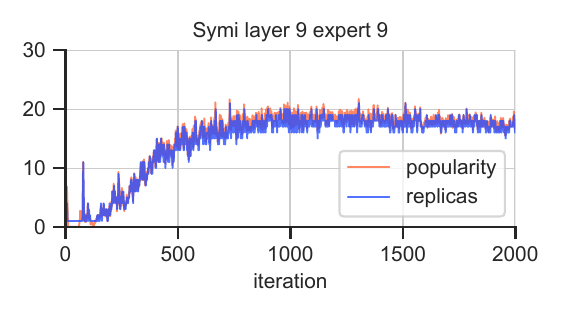}
  \includegraphics[width=0.32\linewidth]{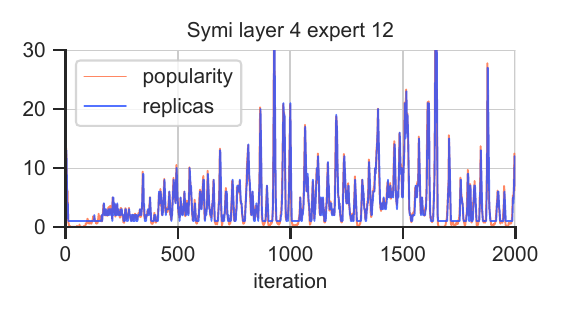}
  \caption{Normalized expert popularity (orange) vs expert replication degree (blue) across different experts in different layers.
  While DeepSpeed allocates a fixed number of expert instances per expert class, \NAME adjusts replication proportionally to expert popularity.
  \NAME’s scheduling policy adapts replication effectively to changing demand -- whether experts lose popularity (left), gain popularity (center), or exhibit highly spiky distributions (right).
  }
  \label{fig:eval_replicas}
\end{figure*}

% !TEX root =paper.tex
\begin{figure}[t]
  \centering
  \includegraphics[width=\linewidth]{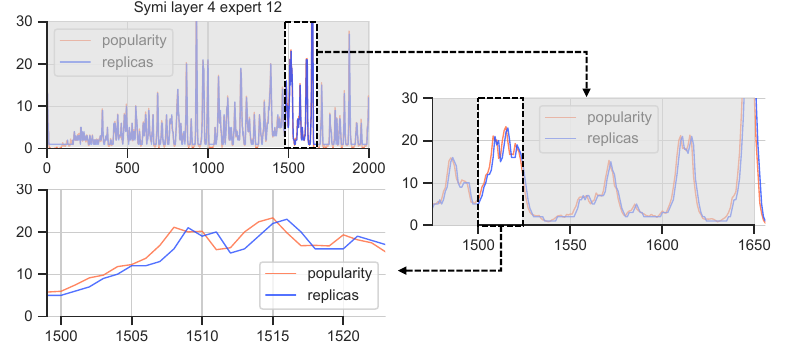}
  \caption{
  Zooming in the expert popularity vs expert replication plot. \NAME's scheduler assigns replicas based on the popularity observed in the previous iteration. This strategy is a good proxy even for very spiky popularity distributions.}
  \label{fig:eval_replicas_zoom}
\end{figure}

Figure~\ref{fig:eval_convergence} shows the training loss of all systems over the course of 2,000 training iterations.
\NAME achieves significantly faster convergence than DeepSpeed throughout training, regardless of the target loss.
For instance, to reach a target loss of 4.0, \NAME requires $28.5\%$ fewer iterations than DeepSpeed.
Also, to reach the same loss \NAME requires $15.6\%$ and $12.1\%$ fewer iterations compared to FlexMoE with low (FlexMoE-100) and medium rebalancing frequencies (FlexMoE-50), respectively.
% also required more training iterations to reach a target accuracy. Specifically, it 
% requires $15.6\%$ (FlexMoE-100) and $12.1\%$ (FlexMoE-50) more training iterations to reach a target loss of 4.0. 
When FlexMoE rebalances every 10 iterations, it requires the same number of iterations to converge as \NAME. 
However, FlexMoE-10's improved convergence comes  at the cost of a significant increase of per-iteration latency (see \S\ref{sec:eval_latency}).

%and $0\%$ (FlexMoE-10) more training iterations to reach a target loss of 4.0.
%While increasing the rebalancing rate of FlexMoE improved the convergence rate at a per-iteration basis, it %comes with extreme latency tradeoffs as we explore in \S\ref{sec:eval_latency}.

%(for reference: Iters to reach loss 4.0, Symi: 1134, FlexMoE-10: 1134 (+0\%), FlexMoE-50: 1290 (+14\%), FlexMoE-100: 1344 (+19\%), DeepSpeed: 1586 (+40\%)}

% Symi: 0.11015494664510073
% FlexMoE-10: 0.19308559290568017
% FlexMoE-50: 0.2867403017679848
% FlexMoE-100: 0.30804553985595695
% DeepSpeed: 0.35179019673665357

To better understand these results, Figure~\ref{fig:eval_survived} shows the fraction of survived (i.e., not dropped) tokens over the course of training.
These results validate two key assumptions.
First, increasing the \textit{frequency} of adaptive replication directly translates to fewer dropped tokens.
In total, \NAME dropped $69\%$, $64\%$, $62\%$ and $43\%$
fewer tokens over the course of training compared to DeepSpeed, FlexMoE-100, FlexMoE-50, and FlexMoE-10, respectively.
Secondly, combined with the results shown in Figure~\ref{fig:eval_convergence}, we observe that dropping fewer tokens allows experts to learn better in each training iteration, directly improving convergence rate on a per-iteration basis.
Thus, Figure~\ref{fig:eval_survived} shows that rebalancing as often as possible directly improves convergence.

%By adaptively replicating expert instances, \NAME dropped \todo{add a specific number} fewer tokens over the course of training compared to DeepSpeed's static replication mechanism.
%By training over more tokens during each iteration, \NAME was able to allow experts to learn better in each training iteration, reinforcing prior works \todo{cite} that show improved convergence as fewer tokens are dropped.
%\askiad{I'll fill exact numbers, but Flex rebalancing on intervals fluctuates. Also cannot go as high as \NAME because of threshold, not rebalancing experts freely/arbitrarily. In general, it is an improvement to DeepSpeed, but does not grasp the full potential of adaptive replication.}

To further understand why more frequent rebalancing drops fewer tokens, Figure~\ref{fig:eval_replicas} compares expert popularity and expert replication for different experts across training.
The top and bottom rows refer to different experts for DeepSpeed and \NAME, respectively.
Meanwhile, the three columns show different commonly observed expert behavior patterns: shrinking popularity, growing popularity, and highly variable popularity.
In the case of DeepSpeed, it is overwhelmingly common that expert popularity largely diverges from the constant expert replication factor.
% the expert popularity does not exactly match the replication factor.
In particular, the middle column shows how very popular experts (as shown in Figure~\ref{fig:motivation_distr}) result in highly-imbalanced replication.
This leads to a large amount of dropped tokens, and as a result, slower convergence.
Meanwhile, the bottom row shows how \NAME is able to adaptively change the replication factor of each expert, under all dynamic behaviors, based on the expert's popularity.

\NAME's expert replication strategy is effective under varied popularity distributions.
Recall that \NAME's Expert Placement Scheduler uses the popularity distribution of the previous iteration.
To validate the efficacy of this technique, Figure~\ref{fig:eval_replicas_zoom} shows a zoomed-in view of the expert popularity and corresponding expert replication during a particularly spiky training interval.
Even in this domain, using the previous iteration as a proxy manages to closely match the expert's dynamic popularity.
% only resulted in \todo{Add a specific number that says how far off we are}.

%\christos{this section is missing a specific statement about prev work on dynamic rebalancing. Since we do it every iteration while others can only afford to do it rarely, how much does thi smatters? You should be able to quote something based on figures 9 and 10, right? FIg 10 shows that if you adopt every (e.g.) 100 iterations, you may completely miss those spikes! THis is a very important point!!!} 
% \input{figures/tex/survived_iterval}
% \askiad{So we also have Figure~\ref{fig:eval_survived_interv}. But convergence doesn't look that amazing -- for this rare case reshuffling every 10 iterations is marginally better (makes sense, hard to explain, harder to explain to reviewer). Do we present the figure or no?}

% !TEX root =paper.tex
\begin{figure}[t]
  \centering
  \includegraphics[width=.49\linewidth]{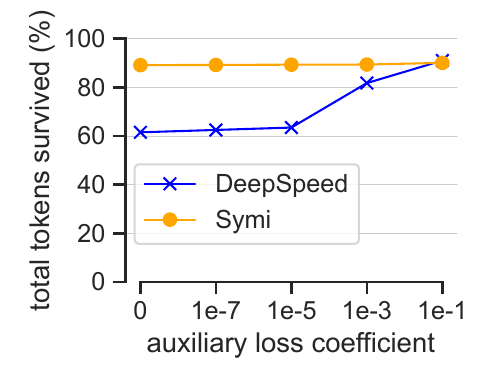}
  \includegraphics[width=.49\linewidth]{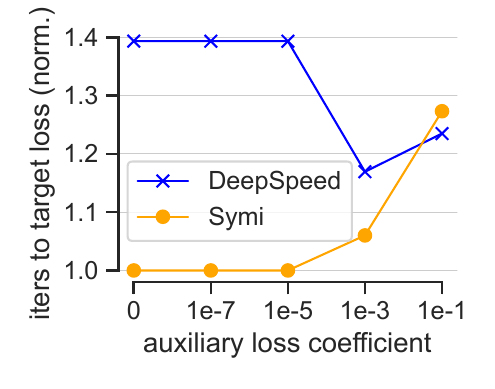}
  \caption{
  % \new{
  Percentage of total survived tokens (left), and normalized iterations needed to reach a target loss (right) for different auxiliary load-balancing loss coefficients.
  In contrast to \NAME, DeepSpeed relies on auxiliary loss involvement to achieve low drop rates and faster convergence.
  % }
  }
  \label{fig:eval_auxloss}
\end{figure}

Adjusting the auxiliary load-balancing loss can help reduce token drops but harms convergence (\S\ref{sec:background_moe}).
In Figure~\ref{fig:eval_auxloss}, we study DeepSpeed's and \NAME's
 behavior under different auxiliary loss coefficients.
DeepSpeed requires a high auxiliary loss coefficient to avoid excessive drops of ${\sim}40\%$ in aggregate. While the lower drop rate helps with convergence, the high auxiliary loss coefficient interferes with the loss objective and with expert routing and bounds the overall benefits in convergence speed. In contrast, the adaptive expert replication allows \NAME to maintain low token drops of ${\sim}10\%$ in aggregate regardless of the coefficient value. \NAME converges fast for all but the very high coefficient values.  
Overall, \NAME removes the need to carefully tune the auxiliary loss to tradeoff convergence speed and system performance,  
turning it into a quality knob rather than a system necessity.
% shifting its role solely to model quality without system tradeoffs.

In summary, these results show that expert popularity changes rapidly, on a per-iteration basis (Figures ~\ref{fig:eval_replicas} and \ref{fig:eval_replicas_zoom}).
Increasing the frequency at which expert placements are adapted according to this shifting popularity is essential to avoiding dropped tokens (Figure~\ref{fig:eval_survived}).
By avoiding these token drops, training systems can drastically improve the convergence rate (Figure~\ref{fig:eval_convergence}).
\NAME adapts expert placements at a \textit{per-iteration} frequency, maximizing the convergence benefits of adaptive replication.

\subsection{Iteration Latency Evaluation}
\label{sec:eval_latency}
% !TEX root =paper.tex
\begin{figure}[t]
  \centering
  \includegraphics[width=\linewidth]{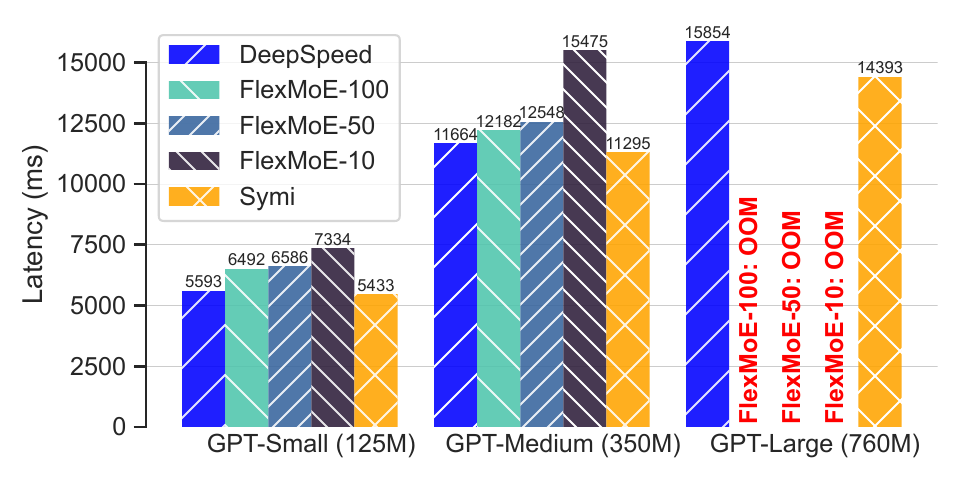}
  \caption{Average iteration latency on different GPT models. \NAME outperforms both the FlexMoE baselines and DeepSpeed. For FlexMoE, average iteration latency increases with rebalancing frequency.}
  \label{fig:eval_latency}
\end{figure}
% !TEX root =paper.tex
\begin{figure}[t]
  \centering
  \includegraphics[width=\linewidth]{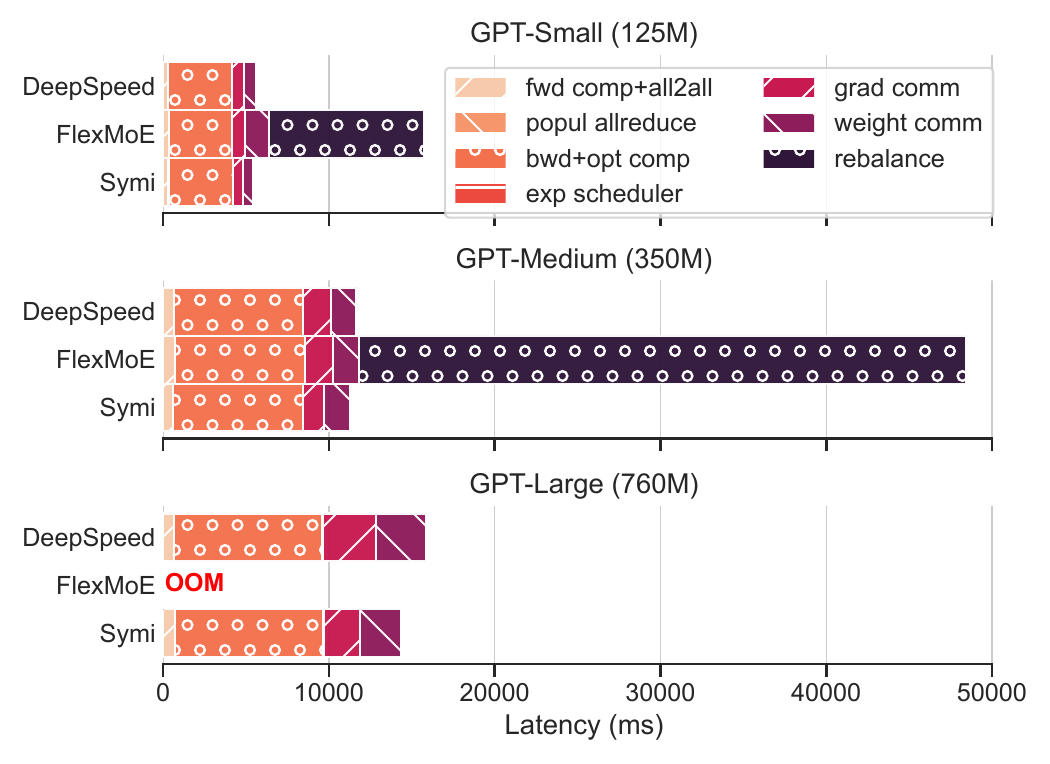}
  \caption{Latency breakdown to the different components of the training iteration. For FlexMoE, we break down iterations where rebalancing occurs. \NAME's newly introduced components add negligible overhead, while in FlexMoE iteration latency spikes.}
  \label{fig:eval_latency_breakdown}
\end{figure}

We validate in practice that \NAME introduces no latency overheads to perform per-iteration adaptive replication.
Figure~\ref{fig:eval_latency} shows the average iteration latency achieved by all systems across the three GPT models. 
\NAME adds no communication overheads over the static DeepSpeed baseline. In fact, it slightly improves iteration latency by $2.8\%$, $3.2\%$, and $9.3\%$ for GPT- Small, Medium, and Large, respectively. 
In contrast, FlexMoE exhibits increasing average iteration latency with higher rebalancing frequencies.
This is the reason we report end-to-end training times for GPT-Small in \S\ref{sec:eval_timetoconv}. 
Also, FlexMoE terminates with an out-of-memory error on GPT-Large as transferring optimizer state requires temporary co-locating current and future state in a given slot. 
% \myzhao{we report GPT-small IIUC. Do we still want to say this?}\christos{the text says GPT-M. Which one is it?} \myzhao{@athinagoras, please check. also, would be good to be consistent with GPT-125 vs GPT-Small, etc.}

%We report the average iteration latency for \NAME and our baselines.
%As running larger models to convergence is increasingly expensive~\cite{arxiv:kaplan_scaling, brown2020language}, we omit GPT Medium and Large from the previous convergence study (\S\ref{sec:eval_latency}).
%Since we implement FlexMoE's scheduling algorithm on top of \NAME with no rebalancing cost, for the current latency study we artificially transfer data worth of an expert's optimizer state over the backend network when an expert migrates to a new rank. \askiad{check how the above sounds}
%\myzhao{moved discussion of this to intro}
%As running larger models to convergence is increasingly expensive~\cite{arxiv:kaplan_scaling, brown2020language}, we omit GTP Medium and Large from the previous convergence study (\S\ref{sec:eval_latency}).
%Figure~\ref{fig:eval_latency} shows that \NAME not only does not add communication overheads over the static DeepSpeed baseline, but also slightly improves latency (by $2.8\%$, $3.2\%$, and $9.3\%$ for GPT Small, Medium, and Large, respectively).
%FlexMoE in contrast exhibits increasingly higher average latency with lower rebalancing frequencies.
%Furthermore, FlexMoE terminates with an out-of-memory error on the 760M model as HBM temporarily collocates the current and the next iteration's experts' optimizer state.

Figure~\ref{fig:eval_latency_breakdown} breaks down each system's latency.
For FlexMoE, we show the average breakdown of a rebalancing iteration. DeepSpeed never rebalances and \NAME rebalances on every iteration.
\NAME's iteration latency improvement over DeepSpeed is attributed to the observed lower communication cost, resulting from our collectives implementation (Section~\ref{sec:comm}).
Additionally, \NAME's new components -- expert popularity all-reduce, expert placement scheduler, metadata update during rebalancing -- introduce negligible overhead, aggregating to only $1.06\%$, $0.82\%$, and $0.70\%$ of the total iteration time for the 125M, 350M, and 760M models, respectively.
% \christos{bound the diff, e.g. within 1\% and make the point that this means that convergence rate is what sets training performance -- so we win. }

Conversely, FlexMoE incurs substantial overhead from optimizer state movement, increasing its rebalancing iterations' latency by $2.46\times$-$4.10\times$. 
While FlexMoE-10 achieved competitive convergence performance to \NAME, frequent rebalancing leads to FlexMoE-10 having ${\sim}35\%$ higher average iteration latency than \NAME.
FlexMoE-50/100  amortize overheads across non-rebalancing iterations. Yet, FlexMoE-50/100 still exhibit higher average iteration latency than \NAME, added to the fact that their infrequent rebalancing strategy compromises convergence.

% Not having to worry about rebalancing overheads is what allows \NAME, to perform arbitrary rebalancing on every iteration. 

\section{Discussion}
\label{sec:discussion}

\textbf{Is \NAME compatible with different offload/parallelism frameworks?}
Our design is compatible with, and can benefit from different offloading and parallelism techniques applied to either experts or the optimizer. 
While this work uses ZeRO-1~\cite{atc21:ren_zero-offload}, ZeRO-2/3 are fully applicable. 
Experts sharded via tensor parallelism/expert-sharding parallelism~\cite{socc22:zhang_spmd, infocom24:pan_parm, ics23:singh_deepspeed-ted} are treated as a single logical expert.
In pipeline parallelism~\cite{narayanan2019pipedream, Chen2023-pipelinemoe}, the optimizer of each layer is uniformly sharded across the ranks of the corresponding pipeline stage.
Scheduling policies that reduce the exposed communication of different forms of parallelism (Section~\ref{sec:related}) can be orthogonally applied to our work.
More broadly, our design decouples model weights from optimizer state but imposes no constraints on how each is internally treated.

\textbf{Is optimizer offload mandatory?}
No. It is a design choice to cleanly separate static components (optimizer state in host DRAM) from dynamic ones (expert weights in GPU HBM).
Offloading state enables usage of model sizes in excess of the storage capacity of the compute complexes. 
Our design 
% \new{
logically
% }
decouples optimizer placement from expert instances but imposes no constraints on where the optimizer must 
% \new{
physically
% }
reside.
In different implementations, the optimizer may be sharded uniformly across accelerator memory 
% \new{
(see Appendix~\ref{apx:nonoffload})
% }
%, offloaded and sharded non-uniformly across different parts of the datacenter serving different experts, 
or may even be stored in disaggregated memory. 
% \new{\NAME gives operators the freedom to choose the configuration that best fits their topology and workload.}
% \NAME unlocks all possible combinations, giving operators the freedom to choose the configuration that best fits their topology and workload.

\textbf{Is \NAME compatible with other replication strategies?}
The recent LLama 4~\cite{llama4} and DeepSeek-V3~\cite{liu2024deepseek} models use a combination of shared and routed experts.
\NAME is applicable to such systems where it would optimize placement for the routed experts.
In general, \NAME alleviates the system bottlenecks in expert placement, allowing AI experts to innovate in learning algorithms without systems limitations. 
%Instead of developers hacking ad-hoc expert placements, \NAME allows the system to determine them.
While we find replicating based on the previous iteration’s popularity effective, \NAME's dynamic replication policy is flexible.
% \NAME supports cost-free, per-iteration re-balancing to arbitrary expert placements, enabling innovation in expert scheduling.
% \NAME can re-balance experts on a per-iteration basis to any new arbitrary placement with no cost, enabling innovation on this front.
The expert scheduler may incorporate prediction, historical statistics, or even disregard popularity alltogether and replicate experts based on expected dataset characteristics. 
%\askiad{changed a bit, how is it now?}
% \christos{I found this answer kind of vague. Most people may conclude that DeepSeek's approach eliminates the need for Symi. Can we say that Symi is applicable to this design, dynamically scaling the non-shared expert?} \christos{if NVL between CPU and GPU are enough, just quicly say this and move on}

\textbf{Is the current hardware architecture optimal for MoE?}
We observed high utilization of the host-to-accelerator path via PCIe and then over the backend network. As optimizer state scales with model size~\cite{sc21:rajbhandari_zero-infinity, cai2025survey}, offloading the optimizer becomes necessary, resulting in increased memory-to-accelerator communication volume and jitter from host memory fetches.
%We expect the need for memory to grow, together with the ability to have hardware support for this memory to be paged and tiered without loss in bandwidth, i.e., latency increases as you go up the tiers, but the net bandwidth does not, will determine the performance of the training system.
% Subsequently, the jitter from host memory fetches will impact operations as the host spends more time with landing user data. 
% As seen in other historical compute trends the need for memory will grow exponentially and the ability to have hardware support for this memory to be paged and tiered without loss in bandwidth i.e. latency increases as you go up the tiers, but the net bandwidth does not, will determine the performance of the training system.
A high-bandwidth and potentially tiered GPU-to-remote-memory interconnect can accelerate MoE systems like \NAME.
\section{Related Work}
\label{sec:related}
SmartMoE~\cite{atc23:zhai_smartmoe} co-locates popular and unpopular experts in the same GPU,
and decides on dynamic rebalancing using a pre-calculated pool of potential expert placements and online dynamic programming.
It does not change the replication of experts and suffers from overheads that prevent arbitrary and frequent expert rebalancing.
SmartMoE is also not applicable to models where the size of an expert crosses the boundaries of HBM.
% and dynamically switch between placement candidates
% via an online dynamic programming approach to co-locate popular and unpopular experts.
% However, it still 
% it does not change the \textit{replication} of experts, efficiently switch experts at iteration granularity.
FasterMoE~\cite{ppopp22:he_fastermoe} instead replicates popular experts to all devices, but requires an additional broadcast after determining the routing decision, introducing latency overheads.
MoESys~\cite{yu2024moesys} dynamically routes experts using hierarchical storage, but also relies on thresholds to balance data transfer costs before applying its mechanisms. 
FlexMoE~\cite{sigmod23:nie_flexmoe} builds upon the notion of adaptive replication found in FasterMoE by \textit{selectively} replicating experts across GPUs according to their popularity distribution.
%This achieves an ideal load-balancing of expert replicas, minimizing the number of dropped tokens.
%However, each re-balancing introduces significant overhead due to copying optimizer state, limiting the frequency that re-balancing can be performed and thus the efficacy of its adaptive replication.
%Section~\ref{sec:background_systems} discussed how SmartMoE~\cite{atc23:zhai_smartmoe}, FasterMoE~\cite{ppopp22:he_fastermoe}, FlexMoE~\cite{sigmod23:nie_flexmoe}, and DeepSpeed~\cite{kdd20:rasley_deepspeed, pmlr22:rajbhandari_deepspeed-moe} cannot meet the distinct requirements presented in Table~\ref{tab:background_tab} for efficient adaptive replication.
%Other recent work focuses on adjacent dimensions of MoE training and inference.

% \new{
Expert choice routing~\cite{zhou2022choice} flips MoE routing so each expert selects its top tokens for better load balance, but is prone to token loss and is not directly applicable to autoregressive decoding.
DeepSeek's auxiliary-loss-free load balancing~\cite{wang2024auxiliary} injects load-based biases directly into router scores instead of altering the loss function.
It uses a tunable update rate and
% has similar tradeoffs as the traditional auxiliary load-balancing loss,
% and additionally requires
a complementary sequence-wise balance loss to avoid extreme imbalances~\cite{liu2024deepseek}. 
Using auxiliary loss or auxiliary-loss-free load balancing is orthogonal to \NAME.
MegaBlocks~\cite{mlsys23:gale_megablocks} unifies variable expert workloads into block-sparse GPU kernels to eliminate drops, but only applies to experts that can fit concurrently in HBM.
MegaBlocks is compatible with, and can benefit \NAME in the intra-device expert domain.
% }

Tutel~\cite{mlsys23:hwang_tutel} leverages data and tensor parallelism for MoE training.
HetuMoE~\cite{arxiv22:nie_hetumoe} supports MoE training on top of the Hetu framework~\cite{scichina22:miao_hetu}.
Janus~\cite{sigcomm23:liu_janus} fetches experts to each GPU as opposed to broadcasting data.
The DeepSpeed~\cite{rasley2020deepspeed, pmlr22:rajbhandari_deepspeed-moe} training framework leverages ZeRO~\cite{rajbhandari2020zero, atc21:ren_zero-offload, sc21:rajbhandari_zero-infinity} techniques to shard and offload each expert's optimizer state across its EDP group.
DeepSpeed does not support adaptive replication.

Recent works, orthogonal to ours, such as ScheMoE~\cite{eurosys24:shi_schemoe}, Parm~\cite{infocom24:pan_parm}, and CCFuser~\cite{ppopp25:wang_ccfuser}, introduce improved collective communication kernels to address the overheads of the all-to-all communication required between experts.
Others, such as Lina~\cite{atc23:li_lina}, PipeMoE~\cite{infocom23:shi_pipemoe}, APTMoE~\cite{sc24:wei_aptmoe}, HiDup~\cite{socc22:zhang_spmd}, FSMoE~\cite{asplos25:pan_fsmoe}, and Lancet~\cite{mlsys24:jiang_lancet}, improve expert sharding and scheduling strategies to better hide communication between experts.
MPMoE~\cite{tpds24:zhang_mpmoe} improves GPU memory efficiency by better managing activations and temporary buffers.
Hexa-MoE~\cite{arxiv25:luo_hexa-moe} optimizes for heterogeneous hardware.

Numerous projects have explored improvements to the expert gating algorithm in MoE models.
GShard~\cite{arxiv:lepikhin_gshard} and Switch Transformers~\cite{jmlr22:fedus_switch-transformers} require a per-expert capacity that drops excess tokens for a given expert, lowering model quality.
BASE Layers~\cite{pmlr21:lewis_baselayers} introduces a linear assignment formulation to improve load balancing between experts.
NetMoE~\cite{iclr25:liu_netmoe} uses an ILP to improve the placement of training samples to reduce all-to-all communication requirements.

\section{Conclusion}
\label{sec:conclusion}
We presented \NAME, an MoE training framework that adaptively replicates expert instances based on training load, on a per-iteration basis.
\NAME's key insight is to decouple each expert's parameters from its optimizer state, allowing \NAME to dynamically adjust the placement of each expert with minimal overhead.
We implemented \NAME on top of DeepSpeed, introducing new components that track expert popularity, schedule expert placements, and communicate across expert instances.
\NAME achieves 25.9\%-30.5\% faster convergence time on GPT compared to DeepSpeed and FlexMoE.
\section*{Acknowledgments}

We thank the anonymous reviewers and our shepherd, T. S. Eugene Ng, whose comments have greatly helped improve this paper.
This research was partly supported by the Stanford Platform Lab and its affiliates, and by ACE, one of the seven centers in JUMP 2.0, a Semiconductor Research Corporation (SRC) program sponsored by DARPA.
Athinagoras Skiadopoulos was partially supported by a Stanford Graduate Fellowship.

%-------------------------------------------------------------------------------
\bibliographystyle{plain}
\bibliography{bibliography}

\clearpage

\appendix
\section{Appendix}

We use the following notation:
\begin{table}[h]
    \centering
    \begin{tabular}{ c | l }
        % \hline
        % Notation & \\
        % \hline \hline
        $N$ & \# nodes in the training cluster \\
        $E$ & \# expert classes \\
        $s$ & \# expert slots per rank \\
        $r$ & \# expert replicas (static baseline) \\
        $r_i$ & \# expert replicas for expert $e_i$ (\NAME) \\
        $BW_{pci}$ & local GPU-CPU interconnect (e.g., PCIe) bandwidth \\
        $BW_{net}$ & cross-node GPU-GPU network (e.g., IB) bandwidth \\
        $G$ & gradients data size for one expert instance \\
        $W$ & weights data size for one expert instance \\
        $T_{G/W}$ & communication cost per rank in each phase \\  
        % \hline
    \end{tabular}
    \caption{Variable Definitions}
    % \label{tab:definitions}
\end{table}

\subsection{Optimal Optimizer Partitioning Strategy}
\label{apx:opt_part}

This appendix proves that partitioning the \NAME optimizer uniformly across all nodes is the most efficient partitioning strategy, given no control over the expert popularity distribution. 
Indeed, we could split the training cluster into $k$ groups, evenly partitioning the optimizer of $\frac{E}{k}$ experts in each \mbox{$\frac{N}{k}$-node} group.

For all experts in the group ($\frac{E}{k}$), the weight/gradient shards ($\frac{X}{N/k}$) are transferred over PCIe ($\frac{1}{BW_{pci}}$).
For all experts instances corresponding to classes outside the group ($\sum_{e_i \in g} (r_i - r_{i|local})$), their gradient/weight shards ($\frac{X}{N/k}$) need to be gathered/updated over the network ($\frac{1}{BW_{net}}$).

For a rank in some group $g$, the communication cost of either $X=G, W$ would be:

\begin{align*}
% T G Symi
T_{X}^{k-part} 
&=~
E/k\frac{X}{N/k} \frac{1}{BW_{pci}}
+
\frac{X}{N/k} \sum_{e_i \in g} (r_i - r_{i|local}) \frac{1}{BW_{net}} \\
&\le~ 
\frac{E}{N} \frac{X}{BW_{pci}}
+
k\frac{(sN - s)}{N} \frac{X}{BW_{net}} \\
\end{align*}
which varies among the $k$ groups and the $\le$ approaches equality for groups containing very popular experts.
The resulting total communication cost, dominated by the highest demand group, increases with $k$.
\NAME ($k=1$) manages to resolve this expensive imbalance as it achieves constant, low communication overhead regardless of expert distribution.

\subsection{\NAME and Static Expert Replication Communication Cost}
\label{apx:comm}

This appendix provides the computation of the formulas presented in \S\ref{sec:design_comm}~(III)

In the static baseline, each expert slot in a rank ($s$) communicates and averages the gradient shards ($\frac{G}{r}$) of the corresponding expert's remote replicas ($r-1$) over the network ($\frac{1}{BW_{net}}$). 
The optimizer then transfers each slot's ($s$) averaged gradient shard ($\frac{G}{r}$) over PCIe ($\frac{1}{BW_{pci}}$). 
After the optimizer step, the updated weights follow the reverse direction: % TODO: introduce the result, e.g. In total, we have:
\begin{align*}
% T G static
T_{G}^{static}
&=~
T_{G|net}^{static}
+
T_{G|local}^{static} \\
&=~
s \frac{(r-1)G}{r} \frac{1}{BW_{net}}
+ 
s \frac{G}{r} \frac{1}{BW_{pci}} \\
&\stackrel{(1)}{\varequals{5pt}}~
s G \Big(
(1 - \frac{E}{sN}) \frac{1}{BW_{net}}
+ 
\frac{E}{sN} \frac{1}{BW_{pci}}
\Big) \\
&=~
% \boxed{
\frac{sN-E}{N} \frac{G}{BW_{net}}
+
\frac{E}{N} \frac{G}{BW_{pci}} \\
% T W static
T_{W}^{static}
&=~
T_{W|local}^{static}
+
T_{W|net}^{static} \\
&=~
s \frac{W}{r} \frac{1}{BW_{pci}}
+
s \frac{(r-1) W}{r} \frac{1}{BW_{net}} \\
&\stackrel{(1)}{\varequals{5pt}}~
s W \Big(
\frac{E}{sN} \frac{1}{BW_{pci}}
+
(1 - \frac{E}{sN}) \frac{1}{BW_{net}}
\Big)  \\
&=~
% \boxed{
\frac{E}{N} \frac{W}{BW_{pci}}
+
\frac{sN-E}{N} \frac{W}{BW_{net}} \\
\end{align*}

In \NAME, each rank needs to communicate and average the gradient shards for the globally partitioned optimizer ($\frac{G}{N}$) of all experts' ($e_i$) remote replicas ($r_i-r_{i|local}$) over the network ($\frac{1}{BW_{net}}$).
The optimizer then transfers all experts' ($E$) averaged gradient shards ($\frac{G}{N}$) over PCIe ($\frac{1}{BW_{pci}}$). 
The updated weights also follow the reverse direction in this case. In total:
\begin{align*}
% T G Symi
T_{G}^{\NAME} 
&=~
T_{G|net}^{\NAME}
+ 
T_{G|local}^{\NAME} \\
&=~
\frac{G}{N} \sum_{e_i} (r_i - r_{i | local}) \frac{1}{BW_{net}}
+
E \frac{G}{N} \frac{1}{BW_{pci}} \\
&\stackrel{(2)}{\varequals{5pt}}~
\frac{G}{N} (sN - s) \frac{1}{BW_{net}}
+
E \frac{G}{N} \frac{1}{BW_{pci}} \\
&=~
% \boxed{
\frac{sN-s}{N} \frac{G}{BW_{net}}
+
\frac{E}{N} \frac{G}{BW_{pci}} \\
% T W Symi
T_{W}^{\NAME}
&=~
T_{W|local}^{\NAME}
+
T_{W|net}^{\NAME} \\
&=~
E \frac{W}{N} \frac{1}{BW_{pci}}
+
\frac{W}{N} \sum_{e_i} (r_i - r_{i | local}) \frac{1}{BW_{net}} \\
&\stackrel{(2)}{\varequals{5pt}}~
E \frac{W}{N} \frac{1}{BW_{pci}}
+
\frac{W}{N} (sN - s) \frac{1}{BW_{net}} \\
&=~
% \boxed{
\frac{E}{N} \frac{W}{BW_{pci}}
+
\frac{sN-s}{N} \frac{W}{BW_{net}} \\
\end{align*}

\subsection{\NAME's Expert Placement Algorithm}
\label{apx:algo_scheduler}
% !TEX root =paper.tex
\begin{codealgorithm}[t]
\caption{The Expert Placement Scheduler's algorithm.}
\begin{lstlisting}[style=mypython]
def compute_placement(popularity, E=exp_classes,
         G=world_size, S=slots_per_rank):
   # initial assignment of instance counts
   goal = (popularity / sum(popularity)) * G * S
   exp_counts = maximum(goal, [1] * E)
   exp_counts = floor(exp_counts)

   # rounding correction
   diff = exp_counts - goal
   while sum(exp_counts) > G * S:
      i = argmax(diff)
      if exp_counts[i] > 1:
         exp_counts[i] -= 1
      diff[i] -= 1
   while sum(exp_counts) < G * S:
      i = argmin(diff)
      exp_counts[i] += 1
      diff[i] += 1

   # assign experts contiguously
   exp_placement = []
   for exp, count in enumerate(exp_counts):
      exp_placement += [exp] * count
   return exp_placement
\end{lstlisting}
\label{alg:scheduler}
\end{codealgorithm}

% \begin{algorithm}[t]
% \caption{The Expert Placement Scheduler's algorithm.}
% \label{alg:scheduler}
% \begin{algorithmic}[1]
% \State $G \gets$ world size
% \State $S \gets$ expert slots per rank

% \State expert\_alloc = \{\}

% \State expert\_distribution = MDS.get\_dist() %\Comment{Get expert distribution from MDS}

% \For{expert, popularity in expert\_distribution}
%     \State expert.assign(1) \Comment{Avoid starvation.}
%     \State expert\_alloc[expert] = max(0, floor(popularity * G * S) - 1)
% \EndFor

% \State instances = []
% \For i, count in enumerate(counts)
%     instances += [i] * count
% \EndFor

% return instances
% \end{algorithmic}
% \end{algorithm}

This appendix shows how the Expert Placement Scheduler assigns experts to slots.
Algorithm~\ref{alg:scheduler} first normalizes the captured expert popularity to the total number of expert slots in the system.
The normalized popularity values are assigned as number of instances to experts, with a minimum of one instance so that all experts remain reachable.
The assigned counts are rounded down, followed by a correction step to ensure that total instances match the available expert slots.
Algorithm~\ref{alg:scheduler} returns an array of expert assignments with same expert classes being contiguous. 
The resulting expert placement favors placement of same-class expert instances within the same rank.

\subsection{\NAME's Gradient Collection Algorithm}
\label{apx:algo_gradgath}
% !TEX root =paper.tex
\begin{codealgorithm}[t]
\caption{\NAME Optimizer's grad collection algorithm.}
\begin{lstlisting}[style=mypython]
def get_source(exp_id, rank):
   if rank in exp_to_rank_map[exp_id]:
      return rank
   candidates = sorted(exp_to_rank_map[exp_id])
   idx = rank % len(candidates)
   return candidates[idx]

def collect_grads():
   # recv/send comm tuples: (rank, partition)
   for exp_id in all_experts:
      recv_grads[exp_id] = (get_source(
            exp_id, self.rank), self.rank)
   for slot, exp_id in local_expert_map.items():
      send_grads[exp_id] = []
      for dst in all_ranks:
         if get_source(exp_id, dst) == self.rank:
            send_grads[exp_id].append((dst, dst))
   set_comm_tuples(recv_grads, send_grads)
\end{lstlisting}
\label{alg:gathergrad}
\end{codealgorithm}

This appendix shows how the \NAME Optimizer assigns ranks during the gradient collection communication.
Algorithm~\ref{alg:gathergrad} assigns a single source rank to a given expert and optimizer destination rank.
The \texttt{\small get\_source} function prioritizes local communication if the expert is local to the optimizer.
Otherwise, \texttt{\small get\_source} round-robins across different expert instances to avoid communication bottlenecks.

\subsection{\NAME With Non-Offloaded Optimizer}
\label{apx:nonoffload}

This appendix shows that \NAME operates with similar benefits when the optimizer is not offloaded, but sharded across HBM memory.
In a static system, each expert is replicated a fixed number of times and its optimizer state is uniformly sharded across the HBM of devices hosting that expert.
In \NAME, the optimizer of all experts is statically and uniformly sharded across the whole HBM domain, while expert instances rebalance dynamically.

To get the equivalent communication cost, we need to set $BW_{pci} \rightarrow \infty$ at the equations in \ref{apx:comm}, giving us:
\begin{align*}
% T G static
T_{G}^{static}
&=~
\frac{sN-E}{N} \frac{G}{BW_{net}} \\
% T W static
T_{W}^{static}
&=~
\frac{sN-E}{N} \frac{W}{BW_{net}} \\
% T G Symi
T_{G}^{\NAME} 
&=~
\frac{sN-s}{N} \frac{G}{BW_{net}} \\
% T W Symi
T_{W}^{\NAME}
&=~
\frac{sN-s}{N} \frac{W}{BW_{net}} \\
\end{align*}

As in \S~\ref{sec:design_comm}, we find the shift in locality increases the communication cost only marginally: 
$$\frac{\Delta T}{T^{static}} = \frac{E-s}{sN - E}$$
which corresponds to just $1.54\%$ in our system example.

% \input{sections/figures/tex/heatmapsd}

%%%%%%%%%%%%%%%%%%%%%%%%%%%%%%%%%%%%%%%%%%%%%%%%%%%%%%%%%%%%%%%%%%%%%%%%%%%%%%%%
\end{document}